\documentclass{article}
\usepackage{amssymb,amsfonts,amsthm,epsfig}
\def\pmb#1{\setbox0=\hbox{$#1$}%
  \kern-.025em\copy0\kern-\wd0
  \kern.05em\copy0\kern-\wd0
  \kern-.025em\raise.0433em\box0}
\def\pmbs#1{\setbox0=\hbox{$\scriptstyle #1$}%
  \kern-.0175em\copy0\kern-\wd0
  \kern.035em\copy0\kern-\wd0
  \kern-.0175em\raise.0303em\box0}
\def\be{\begin{equation}}
\def\ee{\end{equation}}
\def\bea{\begin{eqnarray}}
\def\eea{\end{eqnarray}}
\def\lb{\label}
\def\ct{\cite}
\def\r{\ref}
\def\bi{\bibitem}
\def\lgth{[\,\mbox{length}\,]}

\def\gam{\gamma}

\def\sig{\sigma}
\def\sigp{\sigma_{+}}
\def\sigm{\sigma_{-}}
\def\sigc{\sigma_{\times}}
\def\np{n_{+}}
\def\nm{n_{-}}
\def\nc{n_{\times}}
\def\Sig{\Sigma}
\def\Sigp{\Sigma_{+}}
\def\Sigm{\Sigma_{-}}
\def\Sigc{\Sigma_{\times}}
\def\Np{N_{+}}
\def\Nm{N_{-}}
\def\Nc{N_{\times}}

\def\Om{\Omega}
\def\Oml{\Omega_{\Lambda}}
\def\udot{\dot{u}}
\def\Udot{\dot{U}}
\def\ca{{\cal A}}

\def\cn{{\cal N}}
\def\ce{{\cal E}}
\def\ch{{\cal H}}
\def\ck{\Omega_{k}}
\def\casc{{\cal S}}

\def\p{{\bf e}}
\def\ptl{\partial}
\def\parb{\pmb{\partial}}
\def\e{{\rm e}}

\def\ti{\tilde}
\def\hsp5{\hspace{5mm}}
\newcommand{\sfrac}[2]{{\textstyle{#1\over#2}}}
\def\case#1/#2{\textstyle\frac{#1}{#2}}
\def\apj{{\em Astrophys. J.\/} }
\def\cmp{{\em Commun. Math. Phys.\/} }
\def\cqg{{\em Class. Quantum Grav.\/} }
\def\grg{{\em Gen. Rel. Grav.\/} }
\def\jmp{{\em J. Math. Phys.\/} }

\def\prd{{\em Phys. Rev.\/} D }
\def\prl{{\em Phys. Rev. Lett.\/} }
\def\prs{{\em Proc. R. Soc. Lond. A\/} }
\newcommand{\enl}{\\\hfill\rule{0pt}{0pt}}
\begin{document}
\title{\sc Dynamical systems approach to $G_{2}$ cosmology}
\author{\sc Henk van Elst$^{1}$\thanks{E--mail: {\tt
H.van.Elst@qmul.ac.uk}}\ ,
Claes Uggla$^{2}$\thanks{E--mail: {\tt Claes.Uggla@kau.se}}\ and
John Wainwright$^{3}$\thanks{E--mail:
{\tt jwainwri@math.uwaterloo.ca}}\\
$^{1}${\small\em Astronomy Unit, Queen Mary, University of London,
Mile End Road, London E1 4NS, United Kingdom}\\
$^{2}${\small\em Department of Physics, University of Karlstad,
S--651 88 Karlstad, Sweden}\\
$^{3}${\small\em Department of Applied Mathematics, University of
Waterloo, Waterloo, Ontario, Canada N2L 3G1}}

\date{\normalsize{November 6, 2001}}
\maketitle
\begin{abstract}
In this paper we present a new approach for studying the dynamics
of spatially inhomogeneous cosmological models with one spatial
degree of freedom. By introducing suitable scale-invariant
dependent variables we write the evolution equations of the
Einstein field equations as a system of autonomous partial
differential equations in first-order symmetric hyperbolic format,
whose explicit form depends on the choice of gauge. As a first
application, we show that the asymptotic behaviour near the
cosmological initial singularity can be given a simple geometrical
description in terms of the local past attractor on the boundary of
the scale-invariant dynamical state space. The analysis suggests
the name ``asymptotic silence'' to describe the evolution of the
gravitational field near the cosmological initial singularity.

\end{abstract}
\centerline{\bigskip\noindent PACS number(s): 04.20.-q,
98.80.Hw~\hfill {gr-qc/0107041}}

\section{Introduction}
The simplest cosmological models are the Friedmann--Lema\^{\i}tre
(FL) cosmologies, which describe an expanding Universe that is
exactly spatially homogeneous and spatially isotropic. It is
widely believed that on a sufficiently large spatial scale the
Universe can be described by such a model, at least since the time
of last scattering of primordial photons with unbound electrons.

There are, however, compelling reasons for studying cosmological
models more general than the FL models. Firstly, the observable
part of the Universe is not {\em exactly\/} spatially homogeneous
and isotropic on any spatial scale and so, from a practical point
of view, one is interested in models that are ``close to FL'' in
some appropriate dynamical sense. The usual way to study deviations
from an FL model is to apply {\em linear\/} perturbation
theory. However, it is not known how reliable the linear theory is
and, moreover, in using it one is a priori excluding the
possibility of finding important {\em non-linear\/} effects.
Secondly, it is necessary to consider more general models in order
to investigate the constraints that observations impose on the
geometry of spacetime. Thirdly, it is important to classify all
possible asymptotic states near the cosmological initial
singularity (i.e., near the Planck time) that are permitted by the
Einstein field equations (EFE), with a view to explaining how the
real Universe may have evolved.

For these and other reasons it is of interest to consider a {\em
spacetime symmetry-based hierarchy\/} of cosmological models that
are more general than FL. On the first level above the FL models
are the {\em spatially homogeneous (SH) models\/}, i.e., models
which admit a 3-parameter isometry group acting transitively on
spacelike 3-surfaces, and expand anisotropically. This class has
been studied extensively, and a detailed account of the results
obtained up to 1997 is contained in the book edited by Wainwright
and Ellis (WE) \ct{waiell97}. On the second level of the hierarchy
are cosmological models with two commuting Killing vector fields
(i.e., models which admit a 2-parameter Abelian isometry group
acting transitively on spacelike 2-surfaces), which thus admit one
degree of freedom as regards spatial inhomogeneity. This class of
models, which are referred to briefly as {\em $G_{2}$
cosmologies\/}, are the focus of the present paper. In generalising
from SH cosmologies to $G_{2}$ cosmologies one makes the transition
from ordinary differential equations (ODE) to partial differential
equations (PDE) in two independent variables as regards the
evolution system of the EFE, with the inevitable increase in
mathematical difficulty. For both classes of models one has
available the four standard methods of systematic investigation:
\begin{itemize}
\item[(i)]
derivation and analysis of exact solutions,
\item[(ii)]
approximation methods of a heuristic nature,
\item[(iii)]
numerical simulations and experiments, and
\item[(iv)]
rigorous qualitative analysis.
\end{itemize}
All four methods have been used to study $G_{2}$ cosmologies with
varying degrees of success, subject to significant limitations.

We now give a brief history of $G_{2}$ cosmologies. To the best of
our knowledge, the first development was the study by Gowdy of a
class of solutions of the {\em vacuum\/} EFE with compact space
sections and an Abelian $G_{2}$ isometry group, now called {\em
Gowdy spacetimes\/} \ct{gow71,gow74}. Although vacuum, they can be
regarded as idealised cosmological models because they have a
preferred timelike congruence, start at a curvature singularity,
and either expand indefinitely or recollapse. These solutions could
represent the early stages of the Universe during which the
energy--momentum--stress content is not dynamically significant. As
regards $G_{2}$ cosmologies with a {\em perfect fluid\/} matter
source, the earliest paper was by Liang \ct{lia76}, who used
approximation methods to study the evolution of matter density
fluctuations. A variety of exact perfect fluid $G_{2}$ cosmologies
have been discovered, starting with Wainwright and Goode
\ct{waigoo80}, and more recently by Ruiz and Senovilla
\ct{ruisen92}, Mars and Wolf \ct{marwol97} and Senovilla and Vera
\ct{senver97,senver2001}. Most exact solutions have been derived by
imposing a separability assumption on the metric components, so
that the EFE decouple into two sets of ODE. As regards numerical
simulations, work began in the 1970s (see, e.g., Centrella and
Matzner \ct{cenmat79}). Recent work (e.g., Berger and Moncrief
\ct{bermon93} and Berger and Garfinkle \ct{bergar98}) has focussed
on investigating the nature of the cosmological initial singularity
in Gowdy vacuum spacetimes. Rigorous qualitative analysis has also
focused on the past asymptotic behaviour of these spacetimes,
starting with the paper of Isenberg and Moncrief \ct{isemon90} on
the diagonal subcase. Recent work by Kichenassamy and Rendall
\ct{kicren98} and by Anguige \ct{ang2000b} has considered the Gowdy
vacuum spacetimes with spatial topology ${\mathbb T}^{3}$ and
diagonal perfect fluid $G_{2}$ cosmologies, respectively. We also
refer to Rein \ct{rei96} for related results for a different matter
model.

In summary, almost all of the research using methods (iii) and (iv)
above has focused on the Gowdy vacuum spacetimes. The mathematical
reasons for this choice are twofold: vacuum $G_{2}$ models are much
more tractable than non-vacuum ones, and the assumption of compact
space sections makes numerical simulations easier since it avoids
the problem of boundary conditions at spatial infinity. These works
are nevertheless of considerable physical interest in view of a
conjecture by Belinski\v{\i}, Khalatnikov and Lifshitz (BKL) that
``matter does not matter'' close to the cosmological initial
singularity, i.e., that matter is not dynamically significant in
that epoch (see Lifshitz and Khalatnikov \ct{lifkha63}, p200, and
Belinski\v{\i} {\em et al\/} \ct{bkl70}, p532 and p538). We shall
refer to this conjecture as BKL I.

In this paper we focus on $G_{2}$ cosmologies with perfect fluid
matter content, incorporating vacuum models as an important special
case. The overall goal is to provide a flexible framework
for analysing the evolution of these models in a dynamical systems
context. Our approach, which uses the {\em orthonormal frame
formalism\/},\footnote{See, e.g., MacCallum \ct{mac73}, and for an
extended set of equations, van Elst and Uggla \ct{hveugg97}.} has
three distinctive features:
\begin{itemize}
\item[(i)]
{\em first-order autonomous\/} equation systems,
\item[(ii)]
{\em scale-invariant\/} dependent variables,
\item[(iii)]
evolution equations that form a system of {\em symmetric
hyperbolic\/} PDE.
\end{itemize}
Dynamical formulations employing (i) and (ii) have proved effective
in studying SH cosmologies (see WE \ct{waiell97}, Ch. 5, for
motivation). We expect similar advantages to be gained in the study
of $G_{2}$ cosmologies. In the SH case the scale-invariant
dependent variables are defined by normalisation with the volume
expansion rate of the $G_{3}$--orbits, i.e., the Hubble scalar
$H$. In the present case, however, we define scale-invariant
dependent variables by normalisation with the area expansion rate
of the $G_{2}$--orbits,\footnote{We refer to Hewitt and Wainwright
\ct{hewwai90} for a dynamical formulation of perfect fluid $G_{2}$
cosmologies using Hubble-normalised dependent variables.} in order
to obtain the evolution equations as a system of PDE in first-order
symmetric hyperbolic (FOSH) format. In this way we make available
an additional set of powerful analytical tools, that ensures local
existence, uniqueness and stability of solutions to the Cauchy
initial value problem for $G_{2}$ cosmologies and provides methods
for estimating asymptotic decay rates.\footnote{For details on the
theory of FOSH evolution systems see, e.g., Courant and Hilbert
\ct{couhil62} or Friedrich and Rendall \ct{friren2000}.}  The FOSH
format also provides a natural framework for formulating a concept
of {\em geometrical information propagation\/}, by which we mean
the propagation at finite speeds of jump discontinuities in the
initial data set.\footnote{See, e.g., van Elst {\em et al\/}
\ct{hveetal2000}.}

We now digress briefly to describe some aspects of SH dynamics. The
use of scale-invariant dependent variables to study SH models led
to an important discovery, namely that {\em self-similar
solutions\/} of the EFE play a key r\^{o}le in describing the
dynamics of SH models, in that they can approximate the early,
intermediate and late time behaviour of more general models. We
refer to WE \ct{waiell97}, Ch. 5, for details and other
references. A self-similar solution admits a {\em homothetic vector
field\/}, which, in physical terms, means that as the cosmological
model expands, its physical state differs only by an overall change
in the length scale, i.e., the dynamical properties of the model
are {\em scale-invariant\/}. The above discovery had been
anticipated some years earlier by Eardley \ct{ear74}, who observed
that SH models of Bianchi Type--I, while not self-similar, are {\em
asymptotically self-similar\/}. By this one means that in the
asymptotic regimes, i.e., near the cosmological initial singularity
and at late times, their evolution is approximated by self-similar
models. In other words, these simple models have well-defined
asymptotic regimes that are scale-invariant. In general, however,
SH cosmologies are {\em not\/} asymptotically self-similar. For
example, the well-known Mixmaster models (vacuum solutions of
Bianchi Type--IX; see Ref.~\ct{mis69}) oscillate indefinitely as
the cosmological initial singularity is approached into the past,
and thus do not have a well-defined asymptotic state (see
e.g. Ref.~\ct{bkl70} and WE \ct{waiell97}, Ch. 11). Nevertheless,
as follows from the dynamical systems analysis, the Mixmaster
models are successively approximated by an infinite sequence of
self-similar models (Kasner vacuum solutions) as they evolve into
the past towards the cosmological initial singularity. The
mathematical reason for the above phenomena is that the
self-similar solutions arise as {\em equilibrium points\/} (i.e.,
fixed points) of the evolution equations. These equilibrium points,
in conjunction with the Bianchi classification of the $G_{3}$
isometry group, determine various invariant submanifolds of
increasing generality that provide a hierarchical structure for the
SH dynamical state space. In other words, the self-similar
solutions play a key r\^{o}le as building blocks in determining the
structure of the SH dynamical state space. We anticipate that
self-similar models will play an analogous r\^{o}le in building the
{\em skeleton\/} of the $G_{2}$ dynamical state space. Indeed, the
earlier work of Hewitt and Wainwright \ct{hewwai90} provides some
support for this expectation.

In studying $G_{2}$ cosmologies we expect to make use of
insights into cosmological dynamics obtained from analysing SH
cosmologies, for the following reasons. $G_{2}$ cosmologies can be
regarded as spatially inhomogeneous generalisations of SH models of
all Bianchi isometry group types except Type--VIII and Type--IX,
since, apart from these two cases, the $G_{3}$ admits an Abelian
$G_{2}$ as a subgroup. In the language of dynamical systems the
dynamical state space of SH cosmologies with an Abelian $G_{2}$
subgroup is an invariant submanifold of the {\em dynamical state
space of $G_{2}$ cosmologies\/}. It thus follows that orbits in the
$G_{2}$ dynamical state space that are close to the SH submanifold
will shadow orbits in that submanifold, thereby providing a link
between $G_{2}$ dynamics and SH dynamics. A further link is
provided by the famous conjecture of BKL to the effect that near a
cosmological initial singularity the EFE effectively reduce to ODE,
i.e., the spatial derivatives have a negligible effect on the
dynamics (see Belinski\v{\i} {\em et al\/}
\ct{bkl82}, p656). In this asymptotic regime of near-Planckian
order spacetime curvature it is plausible that SH dynamics will
approximate $G_{2}$ dynamics locally, i.e., along individual
timelines. We shall refer to this conjecture as BKL II.

As indicated above, we expect that SH dynamics will play a
considerable r\^{o}le in determining the dynamics of $G_{2}$
cosmologies, and that analogies with the SH case will be
helpful. In two respects, however, the $G_{2}$ problem differs
considerably from the SH problem. Firstly, at any instant of time,
the state of a $G_{2}$ cosmology is described by a
finite-dimensional dynamical state vector of functions of the
spatial coordinate $x$. In other words, the dynamical state space
of $G_{2}$ cosmologies is a function space and, hence, is {\em
infinite-dimensional\/}. The evolution of a $G_{2}$ cosmology is
thus described by an orbit in this infinite-dimensional dynamical
state space. The second difference lies in the complexity of the
{\em gauge problem\/}, which we now describe. For SH cosmologies
the $G_{3}$ isometry group determines a geometrically preferred
timelike congruence, namely the normal congruence to the
$G_{3}$--orbits, and hence there is a natural choice for the
timelike vector field of the orthonormal frame. The remaining
freedom in the choice of the orthonormal frame is a time-dependent
rotation of the spatial frame vector fields, which we refer to as
the {\em gauge freedom\/}. On the other hand, in a $G_{2}$
cosmology there is a {\em preferred timelike 2-space\/} at each
point that is orthogonal to the $G_{2}$--orbits. Thus there is an
infinite family of geometrically preferred timelike congruences and
the gauge freedom in the choice of the orthonormal frame is
correspondingly more complicated. There is also gauge freedom
associated with the choice of the local coordinates. One of the
goals of the present paper is to systematically discuss various
gauge fixing options that arise for $G_{2}$ cosmologies.

The plan of the paper is as follows. In section~\r{sec:framework}
we derive the equation system that arises from the EFE and the
matter equations. Working in the orthonormal frame formalism, we
adapt the orthonormal frame to the $G_{2}$--orbits and then simply
specialise the general orthonormal frame relations in
Ref.~\ct{hveugg97} to get the desired equation system in
dimensional form. We then introduce the scale-invariant dependent
variables and transform the equation system to dimensionless
form. In section~\r{sec:gauge} we address the gauge problem
and introduce four specific gauge choices, showing that our
approach has the flexibility to incorporate all previous work. In
section~\r{sec:state} we discuss some features of the
infinite-dimensional dynamical state space. In section~\r{sec:past}
we give a simple geometrical representation of the past attractor
as an invariant submanifold on the boundary of the
infinite-dimensional dynamical state space. The nature of the past
attractor illustrates the conjecture BKL II concerning cosmological
initial singularities, and suggests the name ``{\em asymptotic
silence\/}'' to describe the dynamical behaviour of the
gravitational field as one follows a family of timelines into the
past towards the singularity. We conclude in section~\r{sec:concl}
with a discussion of future research directions. Useful
mathematical relations such as expressions for the scale-invariant
components of the Weyl curvature tensor for $G_{2}$ cosmologies and
the propagation laws for the constraint equations have been
gathered in an appendix.

\section{Framework and dynamical equation systems}
\lb{sec:framework}
\subsection{Dimensional equation system}
\lb{subsec:dimeq}
A cosmological model is a triple $\left({\cal M},\,{\bf
g},\,\ti{\bf u}\right)$, where ${\cal M}$ is a 4-dimensional
manifold, ${\bf g}$ a Lorentzian 4-metric of signature $(- + + \,
+)$, and $\ti{\bf u}$ is the matter 4-velocity field. We will
assume that the EFE are satisfied with the matter content being a
{\em perfect fluid\/} with a {\em linear\/} barotropic equation of
state,
\be
\lb{eos}
\ti{p}(\ti{\mu}) = (\gam-1)\,\ti{\mu} \ , \hsp5
1 \leq \gam \leq 2 \ .
\ee
The most important cases physically are radiation ($\gam =
\sfrac{4}{3}$) and dust ($\gam = 1$). We will also include a
non-zero {\em cosmological constant\/} $\Lambda$. Throughout our
work we will employ units characterised by $c = 1 = 8\pi G/c^{2}$.

We assume that an Abelian $G_{2}$ isometry group acts {\em
orthogonally transitively\/} on spacelike 2-surfaces
(cf. Ref.~\ct{car73}), and introduce a group-invariant orthonormal
frame $\{\,\p_{a}\,\}$, with $\p_{2}$ and $\p_{3}$ tangent to the
$G_{2}$--orbits. We regard the frame vector field $\p_{0}$ as
defining a {\em future-directed timelike reference
congruence\/}. Since $\p_{0}$ is orthogonal to the $G_{2}$--orbits,
it is hypersurface orthogonal, and hence is orthogonal to a locally
defined family of spacelike 3-surfaces ${\cal
S}$:$\{t=\mbox{const}\}$. We introduce a set of symmetry-adapted
local coordinates $\{\,t, \,x, \,y, \,z\,\}$ that are tied to the
frame vector fields $\p_{a}$ in the sense that
\be
\lb{framecompos}
\p_{0} = N^{-1}\,\ptl_{t} \ , \hsp5
\p_{1} = e_{1}{}^{1}\,\ptl_{x} \ , \hsp5
\p_{A} = e_{A}{}^{B}\,\ptl_{x^{B}} \ , \hsp5
A, \,B = 2, \,3 \ ,
\ee
where the coefficients are functions of the independent variables
$t$ and $x$ only.\footnote{In the terminology of Arnowitt, Deser
and Misner \ct{adm62}, $N$ is the lapse function, and we have
chosen a zero shift vector field, $N^{i} = 0$ ($i = 1, \,2, \,3$).} 
The only non-zero {\em frame variables\/} are thus given by
\be
N \ , \hsp5 e_{1}{}^{1} \ , \hsp5 e_{A}{}^{B} \ ,
\ee
which yield the following non-zero {\em connection variables\/}:
\be
\alpha, \,\beta, \,a_{1}, \,\np, \,\sigm, \,\nc, \,\sigc,
\,\nm, \,\udot_{1}, \,\Om_{1} \ ;
\ee
their interrelation is given in the appendix. Here we have followed
Ref.~\ct{hveugg97} in doing a $(1+3)$--decomposition of the
connection variables. The variables $\alpha$, $\beta$, $\sigm$ and
$\sigc$ are related to the Hubble volume expansion rate $H$ and the
shear rate $\sig_{\alpha\beta}$ of the timelike reference congruence
$\p_{0}$ according to
\be
\lb{alphabeta}
\alpha := (H-2\sigp) \ , \hsp5 \beta := (H+\sigp) \ ,
\ee
where $\sigp$ is one of the components in the following
decomposition of the symmetric tracefree shear rate tensor
$\sig_{\alpha\beta}$:
\be
\lb{apm}
\sigp := \sfrac{1}{2}\,(\sig_{22}+\sig_{33})
= - \sfrac{1}{2}\,\sig_{11} \ , \hsp5
\sigm := \sfrac{1}{2\sqrt{3}}\,(\sig_{22}-\sig_{33}) \ , \hsp5
\sigc := \sfrac{1}{\sqrt{3}}\,\sig_{23} \ .
\ee
A consequence of this decomposition is that the shear rate scalar
assumes the form $\sig^{2} := \sfrac{1}{2}
(\sig_{\alpha\beta}\sig^{\alpha\beta}) = 3\,(\sigp^{2} + \sigm^{2}
+ \sigc^{2})$. We will use similar decompositions for the electric
and magnetic Weyl curvature variables $E_{\alpha\beta}$ and
$H_{\alpha\beta}$, as given in the appendix. The ``non-null--null''
variables $\alpha$ and $\beta$ (cf. Refs. \ct{uggetal95,ell67})
turn out to be more convenient to use than $H$ and $\sigp$, since
they are naturally adapted to the characteristic structure of the
evolution equations that arise from the Ricci identities when the
latter are applied to the timelike reference congruence $\p_{0}$
(see Ref.~\ct{hveell99}). The variables $a_{1}$, $\np$, $\nc$ and
$\nm$ describe the non-zero components of the purely spatial
commutation functions $a^{\alpha}$ and $n_{\alpha\beta}$, where
\be
\np := \sfrac{1}{2}\,(n_{22}+n_{33}) \ , \hsp5
\nm := \sfrac{1}{2\sqrt{3}}\,(n_{22}-n_{33}) \ , \hsp5
\nc := \sfrac{1}{\sqrt{3}}\,n_{23} \ ,
\ee
(see WE \ct{waiell97} for this type of decomposition of the
spatial commutation functions). Finally, the variable $\udot_{1}$
is the acceleration of the timelike reference congruence $\p_{0}$,
while $\Om_{1}$ represents the rotational freedom of the spatial
frame $\{\,\p_{\alpha}\,\}$ in the $({\bf e}_{2},{\bf
e}_{3})$--plane. Setting $\Om_{1}$ to zero corresponds to the
choice of a Fermi-propagated orthonormal frame $\{\,\p_{a}\,\}$. It
should be pointed out that within the present framework the
dependent variables
\be
\{\,N, \,\udot_{1}, \,\Om_{1}\,\}
\ee
enter the evolution system as freely prescribable {\em gauge source
functions\/} in the sense of Friedrich \ct{fri96}, p1462 (see also
section II.B of Ref. \ct{hveetal2000}).

Since the $G_{2}$ isometry group acts orthogonally transitively,
the 4-velocity vector field $\ti{\bf u}$ of the perfect fluid is
orthogonal to the $G_{2}$--orbits, and hence has the form
\be
\lb{fluid4vel}
\ti{\bf u} = \Gamma\,(\p_{0}+v\,\p_{1}) \ ,
\ee
where the Lorentz factor is $\Gamma := (1-v^{2})^{-1/2}$. It turns
out to be useful to replace the matter energy density $\tilde\mu$
in the fluid rest frame with
\ct{hveugg97}
\be
\mu = \frac{G_{+}}{(1-v^{2})}\,\ti{\mu} \ ,
\ee
where it is convenient to introduce the auxiliary quantities
\be
\lb{gpmdef}
G_{\pm} := 1 \pm (\gam-1)\,v^{2} \ .
\ee
Thus $\mu$, which represents the matter energy density in the rest
frame of $\p_{0}$, and $v$, which describes the matter fluid's
peculiar velocity relative to the same frame, describe, for a given
value of the equation of state parameter $\gam$, the fluid degrees
of freedom.

The orthonormal frame version of the EFE and matter equations as
given in Ref.~\ct{hveugg97}, when specialised to the orthogonally
transitive Abelian $G_{2}$ case with the dependent variables
presented above, takes the following form: \enl

\noindent
{\bf Commutator equations} \nopagebreak

\noindent
{\em Gauge fixing condition\/}:
\be
\lb{gaugefix}
0 = (C_{\udot})_{1} := N^{-1}\,e_{1}{}^{1}\,\ptl_{x}N - \udot_{1}
\ . 
\ee

\noindent
{\em Evolution equations\/}:
\bea
\lb{e11dot}
N^{-1}\,\ptl_{t}e_{1}{}^{1}
& = & -\,\alpha\,e_{1}{}^{1} \\
\lb{e2adot}
N^{-1}\,\ptl_{t}e_{2}{}^{A}
& = & -\,(\beta+\sqrt{3}\,\sigm)\,e_{2}{}^{A}
- (\sqrt{3}\,\sigc+\Om_{1})\,e_{3}{}^{A} \\
\lb{e3adot}
N^{-1}\,\ptl_{t}e_{3}{}^{A}
& = & -\,(\beta-\sqrt{3}\,\sigm)\,e_{3}{}^{A}
- (\sqrt{3}\,\sigc-\Om_{1})\,e_{2}{}^{A} \ .
\eea

\noindent
{\em Constraint equations\/}:
\bea
\lb{e2agrad}
0 & = & (C_{\rm com})^{A}{}_{12} \ := \ (e_{1}{}^{1}\,\ptl_{x}
- a_{1} - \sqrt{3}\,\nc)\,e_{2}{}^{A}
- (\np-\sqrt{3}\,\nm)\,e_{3}{}^{A} \\
\lb{e3agrad}
0 & = & (C_{\rm com})^{A}{}_{31} \ := \ (e_{1}{}^{1}\,\ptl_{x}
- a_{1} + \sqrt{3}\,\nc)\,e_{3}{}^{A}
+ (\np+\sqrt{3}\,\nm)\,e_{2}{}^{A} \ .
\eea

\noindent
{\bf Einstein field equations and Jacobi identities} \nopagebreak

\noindent
{\em Evolution equations\/}:
\bea
\lb{alphadot}
N^{-1}\,\ptl_{t}\alpha & = & -\,\alpha^{2} + \beta^{2}
- 3\,(\sigm^{2}-\nc^{2}+\sigc^{2}-\nm^{2})
- a_{1}^{2} \nonumber \\
& & \hsp5 - \ \sfrac{1}{2}\,\gam\,G_{+}^{-1}\,\mu\,(1-v^{2})
+ (e_{1}{}^{1}\,\ptl_{x}+\udot_{1})\,\udot_{1} \\
\lb{betadot}
N^{-1}\,\ptl_{t}\beta & = & -\,\sfrac{3}{2}\,\beta^{2}
- \sfrac{3}{2}\,(\sigm^{2}+\nc^{2}+\sigc^{2}+\nm^{2})
- \sfrac{1}{2}\,(2\udot_{1}-a_{1})\,a_{1} \nonumber \\
& & \hsp5 - \ \sfrac{1}{2}\,G_{+}^{-1}\,\mu\,[\,(\gam-1)+v^{2}\,]
+ \sfrac{1}{2}\,\Lambda \\
\lb{a1dot}
N^{-1}\,\ptl_{t}a_{1} & = & -\,\beta\,(\udot_{1}+a_{1})
- 3\,(\nc\,\sigm-\nm\,\sigc)
- \sfrac{1}{2}\,\gam\,G_{+}^{-1}\,\mu\,v \\
\lb{npdot}
N^{-1}\,\ptl_{t}\np & = & -\,\alpha\,\np
+ 6\,(\sigm\,\nm+\sigc\,\nc)
- (e_{1}{}^{1}\,\ptl_{x}+\udot_{1})\,\Om_{1} \\
\lb{sigmdot}
N^{-1}\,\ptl_{t}\sigm + e_{1}{}^{1}\,\ptl_{x}\nc
& = & -\,(\alpha+2\beta)\,\sigm
- 2\,\np\,\nm - (\udot_{1}-2a_{1})\,\nc - 2\,\Om_{1}\,\sigc \\
\lb{ncdot}
N^{-1}\,\ptl_{t}\nc + e_{1}{}^{1}\,\ptl_{x}\sigm
& = & -\,\alpha\,\nc + 2\,\sigc\,\np - \udot_{1}\,\sigm
+ 2\,\Om_{1}\,\nm \\
\lb{sigcdot}
N^{-1}\,\ptl_{t}\sigc - e_{1}{}^{1}\,\ptl_{x}\nm
& = &  -\,(\alpha+2\beta)\,\sigc
- 2\,\np\,\nc + (\udot_{1}-2a_{1})\,\nm + 2\,\Om_{1}\,\sigm \\
\lb{nmdot}
N^{-1}\,\ptl_{t}\nm - e_{1}{}^{1}\,\ptl_{x}\sigc
& = & -\,\alpha\,\nm + 2\,\sigm\,\np + \udot_{1}\,\sigc
- 2\,\Om_{1}\,\nc \ .
\eea

\noindent
{\em Constraint equations\/}:
\bea
\lb{gauss}
0 & = & (C_{\rm Gau\ss}) \ := \ 2\,(2\,e_{1}{}^{1}\,\ptl_{x}
-3\,a_{1})\,a_{1}
- 6\,(\nc^{2}+\nm^{2}) + 2\,(2\alpha+\beta)\,\beta
- 6\,(\sigm^{2}+\sigc^{2}) \nonumber \\
& & \hspace{3cm} - \ 2\mu - 2\Lambda \\
\lb{codac}
0 & = & (C_{\rm Codacci})_{1} \ := \ e_{1}{}^{1}\,\ptl_{x}\beta
+ a_{1}\,(\alpha-\beta) - 3\,(\nc\,\sigm-\nm\,\sigc)
- \sfrac{1}{2}\,\gam\,G_{+}^{-1}\,\mu\,v \ .
\eea

\noindent
{\bf Source Bianchi identities (Relativistic Euler equations)}
\nopagebreak

\noindent
{\em Evolution equations\/}:
\bea
\lb{mudot}
\frac{f_{1}}{\mu}\,(N^{-1}\,\ptl_{t}
+ \frac{\gam}{G_{+}}\,v\,e_{1}{}^{1}\,\ptl_{x})\,\mu
+ f_{2}\,e_{1}{}^{1}\,\ptl_{x}v
& = & -\,\frac{\gam}{G_{+}}\,f_{1}\,[\ \alpha\,(1+v^{2})
+ 2\beta + 2\,(\udot_{1}-a_{1})\,v\ ] \\
\lb{vdot}
\frac{f_{2}}{f_{1}}\,\mu\,(N^{-1}\,\ptl_{t}
- \frac{f_{3}}{G_{+}G_{-}}\,v\,e_{1}{}^{1}\,\ptl_{x})\,v
+ f_{2}\,e_{1}{}^{1}\,\ptl_{x}\mu
& = & -\,\frac{f_{2}}{f_{1}G_{-}}\,\mu\,(1-v^{2})\,
[\ (2-\gam)\,\alpha\,v - 2\,(\gam-1)\,\beta\,v \nonumber \\
& & \hspace{20mm} + \ G_{-}\,\udot_{1}
+ 2\,(\gam-1)\,a_{1}\,v^{2}\ ] \ ,
\eea
where
\be
\lb{fdef}
f_{1} := \frac{(\gam-1)}{\gam G_{-}}\,(1-v^{2})^{2} \ , \hsp5
f_{2} := \frac{(\gam-1)}{G_{+}^{2}}\,(1-v^{2})^{2} \ , \hsp5
f_{3} := (3\gam-4)-(\gam-1)\,(4-\gam)\,v^{2} \ .
\ee
Dynamical features exhibited by the dimensional evolution system
for orthogonally transitive $G_{2}$ cosmologies, that are
independent of our later transformation to $\beta$-normalised
scale-invariant dependent variables, are the following:
Eq. (\r{e11dot}) evolves the only dynamically important frame
variable (being part of the {\em metric\/}), Eqs. (\r{alphadot})
and (\r{betadot}) evolve the longitudinal components of the
tensorial {\em expansion\/} rate of the timelike reference
congruence $\p_{0}$, Eqs. (\r{a1dot}) and (\r{npdot}) evolve a
scalar and a non-tensorial {\em spatial connection\/} variable,
respectively, Eqs. (\r{sigmdot}) -- (\r{nmdot}) provide the
propagation laws for (transverse) {\em gravitational waves\/},
while, finally, the relativistic Euler equations (\r{mudot}) and
(\r{vdot}) yield the propagation laws for (longitudinal) {\em
acoustic\/} or {\em pressure waves\/}. Viewing the gauge source
functions $N$, $\udot_{1}$ and $\Om_{1}$ as arbitrarily
prescribable real-valued functions of the independent variables $t$
and $x$, the evolution system is already in FOSH format. The {\em
characteristic propagation velocities\/} $\lambda$ relative to a
family of observers comoving with the timelike reference congruence
$\p_{0}$ are\footnote{On characteristic propagation velocities,
characteristic eigenfields, and related issues see, e.g.,
Refs.~\ct{couhil62}, \ct{hveell99} and \ct{hveetal2000}.}
\be
\lb{charspeed}
\lambda_{1} = 0 \ , \hsp5
\lambda_{2,3} = \pm\,1 \ , \hsp5
\lambda_{4,5} = \frac{(2-\gam)}{G_{-}}\,v
\pm \frac{(1-v^{2})}{G_{-}}\,(\gam-1)^{1/2} \ .
\ee
The right-propagating and left-propagating {\em characteristic
eigenfields\/} associated with the non-zero $\lambda$'s are (for $1
< \gam \leq 2$)\footnote{Here we take the opportunity to correct
for some sign errors in the expressions given in
Refs. \ct{hveell99} and \ct{hveetal2000}.}
\be
\lambda_{2,3}: \ (\sigm\pm\nc) \ , \ (\sigc\mp\nm) \ ,
\hspace{10mm}
\lambda_{4,5}: \ \frac{\mu}{h_{2}(\gam,v)}
\left(h_{2}(\gam,v)\mp h_{1}(\gam,v)
\pm\frac{\gam\,G_{-}^{2}}{(\gam-1)^{1/2}}\,v\right) \ ,
\ee
where $h_{1}(\gam,v)$ and $h_{2}(\gam,v)$ are complicated
expressions of their arguments that for $v = 0$ have the limits
$h_{1} = 0$ and $h_{2} = 1$, respectively. All of the $\lambda$'s
are {\em real-valued\/} in the parameter range $1 \leq
\gam \leq 2$ of Eq. (\r{eos}); this contains the dust case ($\gam
= 1$) and also the stiff fluid case ($\gam = 2$). Note, however,
that the former must be treated in terms of a modified version of
the relativistic Euler equations, since for $\gam = 1$ some of the
coefficients in the principal part of the present version become
zero. In summary, for our first-order dynamical formulation, the
Cauchy initial value problem for the orthogonally transitive
perfect fluid $G_{2}$ cosmologies is {\em well-posed\/} in the
range $1 \leq \gam \leq 2$.\footnote{Clearly well-posedness is lost
for any value of $\gamma$ in the range $0 \leq \gam < 1$
(cf. Ref.~\ct{friren2000}).}

The values of $\lambda_{4,5}$ reflect the anisotropic distortion of
the sound characteristic 3-surfaces relative to the family of
observers comoving with $\p_{0}$. This distortion may be viewed as
a manifestation of the Doppler effect. In the limit $v\rightarrow
0$, the magnitude $|\,\lambda_{4,5}\,|$ reduces to the isentropic
speed of sound, $c_{s} = (\gam-1)^{1/2}$. In the extreme cases
$\gam = 1$ and $\gam = 2$ we obtain $\lambda_{4,5} = v$ and
$\lambda_{4,5} = \pm\,1$, respectively.

Note that in this non-fluid-comoving form [even with the linear
equation of state (\r{eos})] the principal part of the relativistic
Euler equations is highly non-linear. This feature could lead to
the formation of shocks in the fluid dynamical sector of the
evolution system and should be kept in mind in a numerical analysis
of the given equation system. The effective semi-linearity of the
principal part of the gravitational field sector, on the other
hand, is less likely to lead to numerical problems of this
kind. For the latter jump discontinuities can be specified in the
initial data for $\ptl_{x}(\sigm\pm\nc)$ and
$\ptl_{x}(\sigc\mp\nm)$.

The {\em area density\/} $\ca$ of the $G_{2}$--orbits plays a
prominent r\^{o}le for $G_{2}$ cosmologies.\footnote{The symbol we
use to denote the area density, $\ca$, should neither be confused
with the $\beta$-normalised connection variable $A$ to be
introduced below, nor with the index ``$A$'' that takes the values
$2$ and $3$.} It is defined (up to a constant factor) by
\be
\lb{e1}
\ca^{2} := (\xi_{a}\xi^{a})(\eta_{b}\eta^{b})
- (\xi_{a}\eta^{a})^{2} \ ,
\ee
where $\mbox{\boldmath $\xi$}$ and $\mbox{\boldmath $\eta$}$ are
two independent commuting spacelike Killing vector
fields.\footnote{In terms of symmetry-adapted local coordinates
$x^{2} = y$ and $x^{3} = z$ such that $\mbox{\boldmath $\xi$} =
\partial/\partial y$ and $\mbox{\boldmath $\eta$} =
\partial/\partial z$, the area density is given by $\ca =
\sqrt{\det g_{AB}}$, where $g_{AB}$, $A,B = 2,3$, is the metric
induced on the $G_{2}$--orbits.} Expressed in terms of the
coordinate components of the frame vector fields $\p_{A}$ tangent
to the $G_{2}$--orbits this becomes
\be
\ca^{-1} = e_{2}{}^{2}\,e_{3}{}^{3} - e_{2}{}^{3}\,e_{3}{}^{2} \ .
\ee
The {\em key equations\/} for $\ca$, derivable from the commutator
equations (\r{e2adot}) -- (\r{e3agrad}), are given by
\be
\lb{daddim}
N^{-1}\,\frac{\ptl_{t}\ca}{\ca} =  2\beta \ , \hsp5 
e_{1}{}^{1}\,\frac{\ptl_{x}\ca}{\ca} = -\,2a_{1} \ ,
\ee
i.e., $\beta$ is the {\em area expansion rate\/} of the
$G_{2}$--orbits. The frame variables $e_{A}{}^{B}$, which play a
subsidiary r\^{o}le as regards the dynamics of $G_{2}$ cosmologies,
are governed by Eqs. (\r{e2adot}) -- (\r{e3agrad}). Since these
equations are decoupled from the remaining equations, we will not
consider them further.

\subsection{Scale-invariant reduced equation system}
\lb{subsec:dimlesseq}
We will now introduce new {\em dimensionless dependent variables\/}
that are invariant under arbitrary scale transformations. However,
we will {\em not\/} normalise with the Hubble scalar $H$, as is
usually done for SH models (here $H$ is the volume expansion rate
of the $G_{3}$--orbits). Instead we will use the area expansion
rate $\beta$ of the $G_{2}$--orbits, since this leads to
significant mathematical simplifications for the resultant equation
system in both its evolution and constraint part. We thus introduce
$\beta$-normalised frame, connection and curvature variables as
follows:
\bea
\lb{dlframe}
(\,\cn^{-1}, \,E_{1}{}^{1}\,)
& := & (\,N^{-1}, \,e_{1}{}^{1}\,)/\beta \\
(\,\Udot, \,A, \,(1-3\Sigp), \,\Sigm, \,\Nc, \,\Sigc, \,\Nm,
\,\Np, \,R\,)
& := & (\,\udot_{1}, \,a_{1}, \,\alpha, \,\sigm, \,\nc,
\,\sigc, \,\nm, \,\np, \,\Om_{1}\,)/\beta \\
\lb{dlcurv1}
(\,\Om, \,\Oml\,)
& := & (\,\mu, \,\Lambda\,)/(3\beta^{2}) \ .
\eea
Note that we maintain the {\em same\/} notation that was introduced
in WE \ct{waiell97} for the $H$-normalised case. The two different
normalisation procedures are linked through the relation
\be
H = (1-\Sigp)\,\beta \ .
\ee
The above definition of $\Sigp$ in terms of $\alpha$ is motivated
by the relation $\alpha = \beta - 3\sigp$, which follows from
Eqs. (\r{alphabeta}), and is equivalent to defining $\Sigp :=
\sigp/\beta$. As a result of this definition, the
$\beta$-normalised shear rate scalar $\Sig^{2} :=
(\sig_{\alpha\beta}\sig^{\alpha\beta})/(6\beta^{2})$ has the form
$\Sig^{2} = \Sigp^{2} + \Sigm^{2} + \Sigc^{2}$. Note that in the
units we have chosen the matter variable $v$ {\em is\/} already
dimensionless.

The dimensional equation system in subsection \r{subsec:dimeq}
leads to an equation system for the scale-invariant dependent
variables (\r{dlframe}) -- (\r{dlcurv1}). In order to make this
change it is necessary to introduce the time and space rates of
change of the normalisation factor $\beta$. In analogy with
$H$-normalisation (see WE \ct{waiell97}), we define variables $q$
and $r$ by
\bea
\lb{betaq}
\cn^{-1}\,\ptl_{t}\beta & := & -\,(q+1)\,\beta \\
\lb{betar}
0 \ = \ (C_{\beta}) & := & (E_{1}{}^{1}\,\ptl_{x}+r)\,\beta \ .
\eea
Here $q$ plays the r\^{o}le of an ``area deceleration parameter'',
analogous to the usual ``volume deceleration parameter'', while $r$
plays a r\^{o}le analogous to a ``Hubble spatial gradient''. Using
Eqs. (\r{betaq}) and (\r{betar}) and the definitions (\r{dlframe})
-- (\r{dlcurv1}), it is straightforward to transform the
dimensional equation system to a $\beta$-normalised dimensionless
form. A key step is to use the evolution equation (\r{betadot}) for
$\beta$ and the Codacci constraint equation (\r{codac}) to express
$q$ and $r$, as defined above, in terms of the remaining
scale-invariant dependent variables. The key result, which is
essential for casting the scale-invariant evolution system into
FOSH format, is that {\em the expressions for $q$ and $r$ are
purely algebraical\/}. We refer to these equations as {\em the
defining equations for $q$ and $r$\/} [see Eqs. (\r{hdecel}) and
(\r{codac1}) below]. The relation
\be
\lb{betaint}
\cn^{-1}\,\ptl_{t}r - E_{1}{}^{1}\,\ptl_{x}q
= (q+3\Sigp)\,r - (r-\Udot)\,(q+1)
+ (q+1)\,({\cal C}_{\Udot})
\ee
arises as an integrability condition for the decoupled
$\beta$--equations (\r{betaq}) and (\r{betar}). \enl

\noindent
{\bf Scale-invariant equation system} \nopagebreak

\noindent
{\em Evolution system\/}:
\bea
\lb{dle11dot}
\cn^{-1}\,\ptl_{t}E_{1}{}^{1}
& = & (q+3\Sigp)\,E_{1}{}^{1} \\
\lb{dlsigpdot}
3\,\cn^{-1}\,\ptl_{t}\Sigp
& = & -\,3\,(q+3\Sigp)\,(1-\Sigp)
+ 6\,(\Sigp+\Sigm^{2}+\Sigc^{2}) \nonumber \\ 
& & \hsp5 + \ \sfrac{3}{2}\,G_{+}^{-1}\,[\,(3\gam-2)
+(2-\gam)\,v^{2}\,]\,\Om - 3\,\Oml \nonumber \\ 
& & \hsp5 - \ (E_{1}{}^{1}\,\ptl_{x}
-r+\Udot-2A)\,\Udot \\
\lb{dladot}
\cn^{-1}\,\ptl_{t}A
& = & (q+3\Sigp)\,A + (r-\Udot) \\
\lb{dlnpdot}
\cn^{-1}\,\ptl_{t}\Np
& = & (q+3\Sigp)\,\Np + 6\,(\Sigm\,\Nm+\Sigc\,\Nc)
- (E_1{}^1\,\ptl_{x}-r+\Udot)\,R \\
\lb{dlomldot}
\cn^{-1}\,\ptl_{t}\Oml
& = & 2\,(q+1)\,\Oml \\
\lb{dlsigmdot}
\cn^{-1}\,\ptl_{t}\Sigm + E_{1}{}^{1}\,\ptl_{x}\Nc
& = & (q+3\Sigp-2)\,\Sigm - 2\,\Np\,\Nm
+ (r-\Udot+2A)\,\Nc - 2\,R\,\Sigc \\
\lb{dlncdot}
\cn^{-1}\,\ptl_{t}\Nc + E_{1}{}^{1}\,\ptl_{x}\Sigm
& = & (q+3\Sigp)\,\Nc + 2\,\Sigc\,\Np
+ (r-\Udot)\,\Sigm + 2\,R\,\Nm \\
\lb{dlsigcdot}
\cn^{-1}\,\ptl_{t}\Sigc - E_{1}{}^{1}\,\ptl_{x}\Nm
& = & (q+3\Sigp-2)\,\Sigc - 2\,\Np\,\Nc
- (r-\Udot+2A)\,\Nm + 2\,R\,\Sigm \\
\lb{dlnmdot}
\cn^{-1}\,\ptl_{t}\Nm - E_{1}{}^{1}\,\ptl_{x}\Sigc
& = & (q+3\Sigp)\,\Nm + 2\,\Sigm\,\Np
- (r-\Udot)\,\Sigc - 2\,R\,\Nc \ .
\eea
\bea
\lb{dlmudot}
\frac{f_{1}}{\Om}\,(\cn^{-1}\,\ptl_{t}
+ \frac{\gam}{G_{+}}\,v\,E_{1}{}^{1}\,\ptl_{x})\,\Om
+ f_{2}\,E_{1}{}^{1}\,\ptl_{x}v
& = & 2\,\frac{\gam}{G_{+}}\,f_{1}\,[\ \frac{G_{+}}{\gam}\,(q+1)
- \sfrac{1}{2}\,(1-3\Sigp)\,(1+v^{2}) \nonumber \\
& & \hspace{25mm} - \ 1 + (r-\Udot+A)\,v\ ] \\
\lb{dlvdot}
\frac{f_{2}}{f_{1}}\,\Om\,(\cn^{-1}\,\ptl_{t}
- \frac{f_{3}}{G_{+}G_{-}}\,v\,E_{1}{}^{1}\,\ptl_{x})\,v
+ f_{2}\,E_{1}{}^{1}\,\ptl_{x}\Om
& = & 2\,\frac{f_{2}}{f_{1}G_{-}}\,\Om\,(1-v^{2})\,
[\ \frac{(\gam-1)}{\gam}\,(1-v^{2})\,r \\
& & \hspace{15mm} - \ \sfrac{1}{2}\,(2-\gam)\,(1-3\Sigp)\,v
\nonumber \\
& & \hspace{15mm} + \ (\gam-1)\,(1-A\,v)\,v 
- \sfrac{1}{2}\,G_{-}\,\Udot\ ] \ , \nonumber
\eea
where $f_{1}$, $f_{2}$, $f_{3}$ and $G_{\pm}$ are defined by
Eqs. (\r{fdef}) and (\r{gpmdef}), respectively. \enl

\noindent
{\em Defining equations for $q$ and $r$\/}:
\bea
\lb{hdecel}
q & := & \sfrac{1}{2} + \sfrac{1}{2}\,(2\Udot-A)\,A
+ \sfrac{3}{2}\,(\Sigm^{2}+\Nc^{2}+\Sigc^{2}+\Nm^{2})
+ \sfrac{3}{2}\,\frac{(\gam-1) + v^{2}}{G_{+}}\,\Om
- \sfrac{3}{2}\,\Oml \\
\lb{codac1}
r & := & -\,3\,A\,\Sigp - 3\,(\Nc\,\Sigm-\Nm\,\Sigc)
- \sfrac{3}{2}\,\frac{\gam}{G_{+}}\,\Om\,v \ .
\eea
{\em Constraint equations\/}:
\bea
\lb{fried}
0 & = & ({\cal C}_{\rm Gau\ss}) \ = \ \ck - 1 + 2\Sigp
+ \Sigm^{2} + \Sigc^{2} + \Om + \Oml \\
\lb{lamcons}
0 & = & ({\cal C}_{\Lambda}) \ = \ (E_{1}{}^{1}\,\ptl_{x}-2r)\,
\Oml \ ,
\eea
where
\be
\lb{3rscl}
\ck  :=  -\,\sfrac{2}{3}\,(E_{1}{}^{1}\,\ptl_{x}-r)\,A
+ A^{2} + \Nc^{2} + \Nm^{2} \ .
\ee
{\em Gauge fixing condition\/}:
\be
\lb{dlgsfr}
0 = ({\cal C}_{\Udot})
:= \cn^{-1}\,E_{1}{}^{1}\,\ptl_{x}\cn + (r-\Udot) \ .
\ee
\enl

\noindent
{\bf Supplementary equations} \nopagebreak

\noindent
The quantity $(q+3\Sigp)$ occurs frequently in the scale-invariant
equation system. Combining the definition of $q$ given in
Eq. (\r{hdecel}) with the Gau\ss\ contraint equation (\r{fried})
solved for $\Sigp$, one can express this quantity by
\be
\lb{q3Sigp}
(q+3\Sigp) =  2 + (E_{1}{}^{1}\,\ptl_{x} - r + \Udot - 2A)A
- \sfrac{3}{2}\,\frac{(2-\gamma)}{G_{+}}\,(1-v^{2})\,\Om
- 3\Oml \ .
\ee
In terms of our scale-invariant dependent variables, the area
density $\ca$ of the $G_{2}$--orbits satisfies the relations
\be
\lb{dad1}
\ca^{-1}\,\cn^{-1}\,\ptl_t\ca = 2 \ , \hsp5 
\ca^{-1}\,E_{1}{}^{1}\,\ptl_{x}\ca = -\,2A \ .
\ee
Combining the two, the magnitude of the spacetime gradient
$\nabla_{a}\ca$ is
\be
\lb{agrad}
(\nabla_{a}\ca)\,(\nabla^{a}\ca)
= -\,4\beta^{2}\,(1-A^{2})\,\ca^{2} \ ,
\ee
so $\nabla_{a}\ca$ is timelike for $A^{2} < 1$.

\section{Gauge choices}
\lb{sec:gauge}
In this section we discuss the gauge problem.

\subsection{Gauge freedom}
\lb{subsec:gfp}
The scale-invariant equation system in
subsection~\r{subsec:dimlesseq} contains evolution equations for
the dependent variables
\be
\{\,E_{1}{}^{1}, \,\Sigp, \,A, \,\Np, \,\Oml, \,\Sigm, \,\Nc,
\,\Sigc, \,\Nm, \,\Om, \,v\,\} \ ,
\ee
but not for the gauge source functions
\be
\{\,\cn, \,\Udot, \,R\,\} \ ,
\ee
and thus does not uniquely determine the evolution of the $G_{2}$
cosmologies. The reason for this deficiency is that the orthonormal
frame $\{\,\p_{a}\,\}$ and the local coordinates $\{\,t, \,x\,\}$
were not specified uniquely in subsection~\r{subsec:dimeq}. We now
summarise the remaining freedom, which we refer to as the {\em
gauge freedom\/}.

\begin{itemize}
\item[(i)]
Choice of timelike reference congruence $\p_{0}$.

The gauge freedom is a position-dependent {\em boost\/} 
\be
\lb{e3}
\left(\begin{array}{c}
      \hat{\p}_{0} \\
      \hat{\p}_{1}
      \end{array}\right)
= \Gamma\left(\begin{array}{cc}
              1 & w \\
              w & 1
              \end{array}\right)
\left(\begin{array}{c}
       \p_{0} \\
       \p_{1}
       \end{array}\right) \ , \hsp5
\Gamma := \frac{1}{\sqrt{1- w^{2}}} \ , \hsp5
w = w(t,x) \ ,
\ee
in the timelike 2-spaces orthogonal to the $G_{2}$--orbits.

\item[(ii)]
Choice of local time and space coordinates $t$ and $x$.

The gauge freedom is the {\em coordinate reparametrisation\/}
\be
\lb{crepara}
{\hat t} = {\hat t}(t) \ , \hsp5 {\hat x} = {\hat x}(x) \ .
\ee

\item[(iii)]
Choice of spatial frame vector fields $\p_{2}$ and $\p_{3}$.

The gauge freedom is a position-dependent {\em rotation\/}
\be
\lb{23rot}
\left(\begin{array}{c}
      \hat{\p}_{2} \\
      \hat{\p}_{3}
      \end{array}\right)
= \left(\begin{array}{cc}
        \cos\varphi & \sin\varphi \\
       -\sin\varphi & \cos\varphi
        \end{array}\right)
\left(\begin{array}{c}
      \p_{2} \\
      \p_{3}
      \end{array}\right) \ , \hsp5
\varphi = \varphi(t,x) \ ,
\ee
in the spacelike 2-spaces tangent to the $G_{2}$--orbits.

\end{itemize}
We say that (i) and (ii), which refer to the freedom associated
with the preferred timelike 2-space, constitute the {\em temporal
gauge freedom\/}, and that (iii) represents the {\em spatial gauge
freedom\/}.

Table \r{tab1} shows possible ways of fixing the spatial gauge by
requiring one frame vector field or a combination of frame vector
fields to be parallel to a Killing vector field. Each of these sets
of conditions is preserved under evolution and under a boost.
\begin{table}[!htb]
\begin{center}
\begin{tabular}{cc}
\hline
\hline
{\bf Spatial gauge condition} & {\bf Frame vector field} \\
& {\bf parallel to a KVF} \\
\hline
$\Np-\sqrt{3}\Nm = 0 = R+\sqrt{3}\Sigc$ & $\p_{2}$ \\
$\Np+\sqrt{3}\Nm = 0 = R-\sqrt{3}\Sigc$ & $\p_{3}$ \\
$\Np+\sqrt{3}\Nc = 0 = R+\sqrt{3}\Sigm$ & $\p_{2}-\p_{3}$ \\
$\Np-\sqrt{3}\Nc = 0 = R-\sqrt{3}\Sigm$ & $\p_{2}+\p_{3}$ \\
\hline
\hline
\end{tabular}
\end{center}
\caption{Spatial gauge conditions for aligning a combination of the
frame vector fields $\p_{A}$ with a KVF.} 
\lb{tab1}
\end{table}
These choices are essentially all equivalent. We will routinely
make the first choice, namely
\be
\lb{sgfix}
\Np = \sqrt{3}\,\Nm \ , \hsp5 R = -\,\sqrt{3}\,\Sigc \ .
\ee
With this choice the evolution equation (\r{dlnpdot}) becomes
identical to Eq. (\r{dlnmdot}), and thus can be omitted from the
full scale-invariant equation system. Other interesting choices for
fixing the spatial gauge do exist, however, such as a {\em
Fermi-propagated\/} frame, for which $R = 0$.

\subsection{Fixing the temporal gauge}
\lb{subsec:fixtg}
Within the present scale-invariant formulation of the dynamics of
orthogonally transitive $G_{2}$ cosmologies with perfect fluid
matter source we will fix the temporal gauge by adapting the
evolution of the gauge source function $\Udot$ to the following
four geometrical features, listed in order of subsequent
discussion.
\begin{itemize}
\item[(i)]
Adapt the evolution to a family of {\em null\/} characteristic
3-surfaces.
\item[(ii)]
Adapt the evolution to the integral curves determined by the {\em
spacetime gradient of the area density\/} of the $G_{2}$--orbits,
$\nabla_{a}\ca$.
\item[(iii)]
Adapt the evolution to the family of fluid {\em sound\/}
characteristic 3-surfaces.
\item[(iv)]
Adapt the evolution to {\em zero-velocity\/}
characteristic 3-surfaces associated with a family of
freely-falling observers.
\end{itemize}
The idea is to specialise $\p_{0}$ in such a way that either
$\cn^{-1}\,\ptl_{t}\Udot$ or $\Udot$ itself is determined in terms
of the other dependent variables. Then $\cn$ is determined from
Eq. (\r{dlgsfr}) up to an arbitrary dimensionless multiplicative
function $f(t)$. We then use a reparametrisation of $t$ to choose
\be
f(t) = \e^{Ct} \ ,
\ee
where $C$ is an arbitrary constant. This coordinate choice leads to
an autonomous differential equation for $\cn$ which we include in
the evolution system, giving a fully determined {\em autonomous\/}
scale-invariant equation system.

It should be pointed out that apart from the above four choices of
temporal gauge other interesting possibilities such as, e.g., a
constant area expansion rate gauge, where $r = 0
\Leftrightarrow \beta = \beta(t)$, do exist.

\subsubsection{Null cone gauge}
\lb{subsubsec:ncg}
The first choice of gauge, which we call the {\em null cone
gauge\/}, is motivated by the identity
\be
\lb{dlid1}
\cn^{-1}\,\ptl_{t}(r-\Udot) - E_{1}{}^{1}\,\ptl_{x}(q+3\Sigp)
= -\,\cn^{-1}\,E_{1}{}^{1}\,
\ptl_{t}\ptl_{x}[\,\ln(\cn E_{1}{}^{1})\,]
+ \cn^{-1}\,\ptl_{t}({\cal C}_{\Udot}) \ ,
\ee
which follows from combining Eqs. (\r{dlgsfr}) and
(\r{dle11dot}). It suggests that we impose the condition
\be
\lb{ncgc}
0 = \cn^{-1}\,\ptl_{t}(r-\Udot) - E_{1}{}^{1}\,\ptl_{x}(q+3\Sigp)
\ee
on $\cn^{-1}\,\ptl_{t}\Udot$. It follows immediately from these two
relations and Eq. (\r{betaint}) that
\be
\lb{ncgn}
\cn = \frac{f(t)g(x)}{E_{1}{}^{1}}
\ee
and
\be
\lb{ncgudotdot}
\sfrac{1}{3}\,\cn^{-1}\,\ptl_{t}\Udot
+ E_{1}{}^{1}\,\ptl_{x}\Sigp
= \sfrac{1}{3}\,(q+3\Sigp)\,\Udot
- \sfrac{1}{3}\,(r-\Udot)\,(1-3\Sigp) \ ,
\ee
provided that the gauge fixing condition (\r{dlgsfr}) propagates
along $\p_{0}$ according to Eq. (\r{gfcpropnc}) in the appendix. We
now use the $t$-reparametrisation (\r{crepara}) to set $f(t) =
\e^{C_{\rm nc}t}$, where $C_{\rm nc}$ is a constant. Equation
(\r{ncgn}) then gives
\be
\lb{ncgc1}
\cn E_{1}{}^{1} = \e^{C_{\rm nc}t}\,g(x) \ .
\ee
On differentiating Eq. (\r{ncgc1}) and using Eq. (\r{dle11dot}), we
obtain an evolution equation for $\cn$ that reads
\be
\lb{ncgndot}
\cn^{-1}\,\ptl_{t}\cn = -\,(q+3\Sigp)\,\cn + C_{\rm nc} \ .
\ee
Note that in the null cone gauge Eqs. (\r{dlsigpdot}) and
(\r{ncgudotdot}) form the $(\Sigp,\Udot)$--branch of an autonomous
evolution system in FOSH format. The associated characteristic
propagation velocities are $\lambda = \pm\,1$.

Choosing the null cone gauge permits one to introduce the familiar
{\em conformal coordinates\/} $\{\,t,\,x\,\}$ in the timelike
2-spaces orthogonal to the $G_{2}$--orbits, although we do not find
it convenient to make this choice in general. Referring to
Eq.~(\r{ncgn}), one can use the coordinate reparametrisation
(\r{crepara}) to set $f(t) = 1$ and $g(x) = 1$, so that $\cn
E_{1}{}^{1} = 1$. It follows from Eq. (\r{dlframe}) that
$Ne_{1}{}^{1} = 1$, which implies, using Eqs. (\r{framecompos}),
that the line element in the timelike 2-spaces orthogonal to the
$G_{2}$--orbits has the form
\be
{}^{(2)}\!{\rm d}s^{2} = N^{2}\,(-\,{\rm d}t^{2}+{\rm d}x^{2}) \ .
\ee
Conformal coordinates have been frequently used in the analytical
study of vacuum $G_{2}$ cosmologies, and in the derivation of exact
solutions, both for vacuum and for perfect fluid models. Selected
references from the literature are Gowdy \ct{gow71,gow74}, Liang
\ct{lia76}, Isenberg and Moncrief \ct{isemon90}, H\"{u}bner
\ct{hue98}, Kichenassamy and Rendall \ct{kicren98}, Senovilla and
Vera \ct{senver98} and Anguige \ct{ang2000b}.

\subsubsection{Area gauges}
\lb{subsubsec:areag}
The {\em separable area gauge\/} is determined by imposing the
condition
\be
\lb{areagc}
0 = (r-\Udot) \ ,
\ee
which determines $\Udot$ algebraically through Eq. (\r{codac1}).
There is thus no need to determine an evolution equation for
$\Udot$. It follows immediately from the gauge fixing condition
(\r{dlgsfr}) that $\cn = f(t)$. We now use the
$t$-reparametrisation (\r{crepara}) to set $f(t) =
\cn_{0}$, a constant, i.e.,
\be
\lb{areagn}
\cn = \cn_{0} \ .
\ee
In this case the evolution equation for $\cn$ is {\em trivial\/},
i.e.,
\be
\cn^{-1}\,\ptl_{t}\cn = 0 \ .
\ee
It follows from Eqs. (\r{dad1}) and (\r{areagn}) that the area
density has the form
\be
\lb{areadens}
\ca = \ell_{0}^{2}\,\e^{2\cn_{0}t}\,m(x) \ ,
\ee
where here and throughout $\ell_{0}$ denotes the unit of the
physical dimension $\lgth$, and $m(x)$ is a positive function of
$x$. The gauge fixing condition (\r{areagc}) propagates along
$\p_{0}$ according to Eq. (\r{gfcpropsa}) in the appendix subject
to an auxiliary equation for $\cn_{0}^{-1}\,\ptl_{t}\Udot$. Note
that the separable area gauge does not in general yield an
evolution system in FOSH format.

For the class of $G_{2}$ cosmologies in which the spacetime
gradient $\nabla_{a}\ca$ is {\em timelike\/}, we can strengthen the
separable area gauge condition (\r{areagc}) by requiring in
addition that
\be
\lb{tlareagc}
A = 0 \ ,
\ee
which we achieve by choosing $\p_{0}$ to be parallel to
$\nabla_{a}\ca$. It follows from Eq. (\r{dad1}) that $\ptl_{x}\ca =
0$, and Eq. (\r{areadens}) reduces to
\be
\ca = \ell_{0}^{2}\,\e^{2\cn_{0}t} \ .
\ee
This defines the so-called area time coordinate. Observe that
condition (\r{tlareagc}) is invariant, by virtue of
Eqs. (\r{dladot}) and (\r{areagc}). We shall refer to the gauge
choices (\r{areagc}) and (\r{tlareagc}) as the {\em timelike area
gauge\/}. We note that in this case, with $\Sigp$ and $\Udot$
algebraically determined in terms of the other scale-invariant
dependent variables from, respectively, the Gau\ss\ constraint
equation (\r{fried}) and Eqs. (\r{areagc}) and (\r{codac1}), the
evolution system becomes unconstrained when $\Oml = 0$ and does
assume FOSH format. We give the resultant equation system in
subsection \r{subsec:gst}. Selected references from the literature
using the timelike area gauge are Berger and Moncrief
\ct{bermon93}, Hern and Stewart \ct{herste98} and Rendall and
Weaver \ct{renwea2001}.

\subsubsection{Fluid-comoving gauge}
\lb{subsubsec:fcomg}
The {\em fluid-comoving gauge\/} is determined by choosing $\p_{0}$
to be equal to the fluid 4-velocity field $\ti{\bf u}$. By virtue
of Eq. (\r{fluid4vel}), this choice is equivalent to imposing the
condition
\be
\lb{fcgc}
0 = v \ .
\ee
The evolution equations (\r{dlmudot}) and (\r{dlvdot}) for $\Om$
and $v$ now reduce to
\bea
\lb{cmomdot}
\cn^{-1}\,\ptl_{t}\Om
& = & 2\,[\ (q+3\Sigp-2) + \sfrac{3}{2}\,(2-\gam)\,(1-\Sigp)\ ]
\ \Om \\
\lb{cmmomcons}
0 & = & ({\cal C}_{v}) \ := \ [\,(\gam-1)\,
(E_{1}{}^{1}\,\ptl_{x}-2r) + \gam\,\Udot\,]\ \Om \ .
\eea
The evolution equation for $\Udot$ (which is now identified with
the $\beta$-normalised fluid acceleration) results from demanding
that the {\em new\/} constraint equation (\r{cmmomcons}) propagates
along $\ti{\bf u}$. This leads to
\be
\lb{fcgudotdot}
\sfrac{1}{3}\,\cn^{-1}\,\ptl_{t}\Udot
+ (\gam-1)\,E_{1}{}^{1}\,\ptl_{x}\Sigp
= \sfrac{1}{3}\,(q+3\Sigp)\,\Udot
- (\gam-1)\,(1-\Sigp)\,(r-\Udot) \ .
\ee
It then follows that the gauge fixing condition (\ref{dlgsfr})
propagates along $\ti{\bf u}$ according to Eq. (\r{gfcpropco}) in
the appendix. Furthermore, it follows from Eqs. (\r{cmmomcons}) and
(\ref{dlgsfr}) that\footnote{This equation is a scale-invariant
form of the well-known dimensional relation $N = a(t)\,
\mu^{-(\gam-1)/\gam}$ for perfect fluid models with
equation of state $p(\mu) = (\gam-1)\,\mu$ in fluid-comoving gauge;
see, e.g., Ref.~\ct{syn37}.}
\be
\lb{fcgn}
\cn = f(t)\,(\ell_{0}\beta)\left(\ell_{0}^{2}\beta^{2}\,\Omega
\right)^{-(\gam-1)/\gam} \ .
\ee
We now use the $t$-reparametrisation (\r{crepara}) to set $f(t) =
\e^{C_{\rm fc}t}$. On differentiating Eq. (\r{fcgn}) and using
Eqs. (\r{betaq}) and (\r{cmomdot}), we obtain an evolution equation
for $\cn$ that reads
\be
\lb{fcgndot}
\cn^{-1}\,\ptl_{t}\cn
= -\,[\ (q+3\Sigp-2) + 3\,(2-\gam)\,(1-\Sigp)\ ]\ \cn
+ C_{\rm fc} \ .
\ee
In order to obtain an evolution system in FOSH format, we need to
multiply Eq. (\r{dlsigpdot}) by a factor of $(\gam-1)$ and write
it in the form
\bea
\lb{fcgsigpdot}
(\gam-1)\,(3\,\cn^{-1}\,\ptl_{t}\Sigp + E_{1}{}^{1}\,\ptl_{x}\Udot)
& = & -\,(\gam-1)\,[\ (q+3\Sigp)\,(1-3\Sigp) + 2q
- 6\,(\Sig_{-}^{2}+\Sig_{\times}^{2})
\nonumber \\ 
& & \hspace{15mm} - \ (r-\Udot+2A)\,\Udot
- \sfrac{3}{2}\,(3\gam-2)\,\Om + 3\Oml\ ] \ .
\eea
When we adjoin Eqs. (\r{fcgudotdot}), (\r{fcgndot}) and
(\r{fcgsigpdot}) to the full scale-invariant equation system, with
the fluid evolution equations (\r{dlmudot}) and (\r{dlvdot})
replaced by Eq. (\r{cmomdot}) and the new constraint equation
(\r{cmmomcons}), and Eq. (\r{dlsigpdot}) replaced by
Eq. (\r{fcgsigpdot}), we obtain a new evolution system that has
FOSH format, with the fluid dynamical sector being shifted from the
$(\Om,v)$--branch to the $(\Sigp,\Udot)$--branch, with
characteristic velocities given by $\lambda_{4,5} =
\pm\,(\gam-1)^{1/2}$. The family of sound characteristic 3-surfaces
thus becomes symmetrically embedded inside the family of null
characteristic 3-surfaces. In the case of dust, $\gam = 1
\Leftrightarrow \Udot = 0$, Eq. (\r{fcgsigpdot}) does apply {\em
without\/} the common factor $(\gam-1)$.

When doing numerical experiments in the present framework, the
fluid-comoving gauge will have the advantage that, in view of the
{\em linear\/} equation of state (\r{eos}), the effective
semi-linearity of the principal part of Eqs. (\r{fcgsigpdot}) and
(\r{fcgudotdot}) will prevent the development of shocks in the
fluid dynamical sector of the evolution system. Hence, only the
propagation of so-called contact discontinuities is possible in
both the gravitational field and the fluid dynamical
sectors. Another advantage of this gauge is that it makes direct
physical interpretation possible in terms of kinematical fluid
quantities.

Examples of references employing the fluid-comoving gauge are
Eardley {\em et al\/} \ct{earetal72}, Wainwright and Goode
\ct{waigoo80} and Ruiz and Senovilla \ct{ruisen92}.

\subsubsection{Synchronous gauge}
\lb{subsubsec:syncg}
The {\em synchronous gauge\/} is determined by choosing $\p_{0}$ to
be a timelike reference congruence that is {\em geodesic\/}, i.e.,
we impose the condition
\be
\lb{syncgc}
0 = \Udot \ .
\ee
It follows from the gauge fixing condition (\r{dlgsfr}) and
Eq. (\r{betar}) that
\be
\lb{syncgn}
\cn = f(t)\,(\ell_{0}\beta) \ .
\ee
We now use the $t$-reparametrisation (\r{crepara}) to set $f(t) =
\e^{C_{\rm sync}t}$. On differentiating Eq. (\r{syncgn}) and using
Eq. (\r{betaq}), we obtain an evolution equation for $\cn$ that
reads
\be
\lb{syncndot}
\cn^{-1}\,\ptl_{t}\cn = -\,(q+1)\,\cn + C_{\rm sync} \ .
\ee
The gauge fixing condition (\r{dlgsfr}) presently propagates along
$\p_{0}$ according to Eq. (\r{gfcpropsync}) in the appendix. When
we adjoin Eq. (\r{syncndot}) to the full scale-invariant evolution
system, simplified using Eq. (\r{syncgc}), we again obtain FOSH
format. In a more general context, the synchronous gauge has been
made prominent in particular by the work of BKL \ct{bkl70,bkl82}.

\section{Scale-invariant dynamical state space}
\lb{sec:state}
The description of the dynamics of $G_{2}$ cosmologies is
complicated by the fact that the scale-invariant dynamical state
vector, and hence the structure of the scale-invariant dynamical
state space, depends on the choice of gauge. In this section we
discuss this issue, and we explain which properties of the
scale-invariant equation system and of the dynamical state space
are independent of the choice of gauge. For simplicity, in the
present discussion we set the cosmological constant to zero, $\Oml
= 0$.

\subsection{Overview}
\lb{subsec:ov}
We assume that the spatial gauge has been fixed according to
Eq. (\r{sgfix}). Once we choose a specific temporal gauge, the
equation system derived in subsection \r{subsec:dimlesseq} gives an
explicit set of evolution and constraint equations for a
finite-dimensional dynamical state vector ${\bf X}$. These
equations can be written concisely in the following form, where the
FOSH nature of the evolution part is indicated by the fact that the
coefficient matrices ${\bf A}({\bf X})$ and ${\bf B}({\bf X})$ are
{\em symmetric\/}, with ${\bf A}({\bf X})$ being {\em positive
definite\/}. \enl

\noindent
{\em Evolution system\/}:
\be
\lb{ovevol}
{\bf A}({\bf X})\,\parb_{0}{\bf X}
+ {\bf B}({\bf X})\,\parb_{1}{\bf X}
= {\bf F}({\bf X}) \ ,
\ee
and \enl

\noindent
{\em Constraint equations\/}:
\be
\lb{ovconstr}
0 = {\bf\cal C}({\bf X},\parb_{1}{\bf X}) \ ,
\ee
where $\parb_{0} := \cn^{-1}\,\ptl_{t}$ and $\parb_{1} :=
E_{1}{}^{1}\,\ptl_{x}$. The dynamical state vector ${\bf X}$
depends on the choice of temporal gauge as follows:
\be
{\bf X} = {\bf X}_{\rm g} \oplus {\bf X}_{\rm w} \ ,
\ee
where ${\bf X}_{\rm g}$ is the {\em temporal gauge-dependent\/}
part, and
\be
\lb{gravdyndegfree}
{\bf X}_{\rm w} = (\,\Sigm, \,\Nc, \,\Sigc, \,\Nm\,)^{T} \ ,
\ee
the {\em temporal gauge-independent\/} part describing the
dynamical degrees of freedom in the gravitational field. The latter
amount to four arbitrary real-valued functions of $x$ one can
specify initially.
\begin{table}[!htb]
\begin{center}
\begin{tabular}{ccccc}
\hline
\hline
{\bf Temporal} & ${\bf X}_{\rm g}$ & {\bf Constraint} &
{\bf Initially freely} & {\bf No. of gauge} \\
{\bf gauge} &  &
{\bf equations} & {\bf specifiable} & {\bf degrees} \\
 & & & {\bf functions} & {\bf of freedom} \\
\hline
Fluid-comoving & $(\cn, E_{1}{}^{1}, \Sigp, \Udot, A, \Om\,)^{T}$ &
$({\cal C}_{\Udot})$, $({\cal C}_{\rm Gau\ss})$,
$({\cal C}_{v})$ & $(E_{1}{}^{1}, A, \Om\,)^{T}$ & $1$
\\
\hline
Timelike area & $(E_{1}{}^{1}, \Om, v)^{T}$ & none &
$(E_{1}{}^{1}, \Om, v\,)^{T}$ & $1$ \\
\hline
Separable area & $(E_{1}{}^{1}, \Sigp, A, \Om, v\,)^{T}$
& $({\cal C}_{\rm Gau\ss})$ & $(E_{1}{}^{1}, A, \Om, v)^{T}$ &
$2$ \\
\hline
Synchronous & $(\cn, E_{1}{}^{1}, \Sigp, A, \Om, v)^{T}$ &
$({\cal C}_{\Udot})$, $({\cal C}_{\rm Gau\ss})$ &
$(E_{1}{}^{1}, A, \Om, v)^{T}$ & $2$ \\
\hline
Null cone & $(\cn, E_{1}{}^{1}, \Sigp, \Udot, A, \Om, v)^{T}$ &
$({\cal C}_{\Udot})$, $({\cal C}_{\rm Gau\ss})$ & $(E_{1}{}^{1},
\Udot, A, \Om, v)^{T}$ & $3$ \\
\hline
\hline
\end{tabular}
\end{center}
\caption{Form of ${\bf X}_{\rm g}$ and number of gauge degrees of
freedom for different temporal gauge choices.}
\lb{tab2}
\end{table}

\noindent
In Tab.~\r{tab2} we show the number of gauge degrees of freedom
that remain after choosing a specific temporal gauge. This number
is arrived at as follows:
\begin{eqnarray*}
\mbox{No.(initially freely specifiable functions)}
& = & \mbox{Dim}({\bf X}_{\rm g}) - \mbox{No.(constraint equations)}
\\
\mbox{No.(gauge degrees of freedom)}
& = & \mbox{No.(initially freely specifiable functions)} - 2 \ .
\end{eqnarray*}
We now discuss how the remaining gauge degrees of freedom arise.
Once a temporal gauge has been chosen, each set of initial
conditions
$$
{}_{0}\!{\bf X} = {\bf X}(t_{0},x)
$$
that satisfies the constraint equations determines a unique
solution of the evolution equations. Because of the remaining gauge
freedom, different initial conditions do not necessarily lead to
physically distinct solutions. We can use the remaining gauge
freedom to simplify the initial conditions as follows. Firstly, the
$x$-reparametrisation (\r{crepara}) can be used to set
$$
E_{1}{}^{1}(t_0,x) = 1 \quad \mbox{for all}\ x
$$
in each gauge, thereby eliminating one gauge degree of freedom. At
this stage there is no remaining gauge freedom in the
fluid-comoving and timelike area gauges. The remaining gauge
freedom in the three other temporal gauges is a boost (\r{e3}) with
velocity $w = w(t,x)$ that preserves the appropriate gauge fixing
condition. The requirement that the gauge fixing condition be
preserved leads to a propagation equation for $w(t,x)$, of {\em
first order in time\/} for the separable area and synchronous
gauges and of {\em second order in time\/} for the null cone
gauge. Thus, for the first two gauges one has the freedom to
perform a boost at the initial time with $w(t_{0},x)$ arbitrary,
while in the null cone gauge both $w(t_{0},x)$ and $\ptl_{t}
w(t_{0},x)$ can be chosen arbitrarily. One can use this restricted
boost with $w(t_0,x)$ arbitrary to set $v(t_{0},x) = 0$, and in the
case of the null cone gauge one can use the arbitrary function
$\ptl_{t} w(t_{0},x)$ to also set $\Udot(t_{0},x) = 0$. In other
words, in the separable area and synchronous gauges one can,
without loss of generality, use the initial condition $v(t_{0},x) =
0$, and in the null cone gauge one can likewise, without loss of
generality, use the initial conditions $v(t_{0},x) = 0$ and
$\Udot(t_{0},x) = 0$.  Thus, in each gauge, ${\bf X}_{\rm g}$ is
specified initially by giving two arbitrary real-valued functions
of $x$, which, with ${\bf X}_{\rm w}$ in Eq. (\r{gravdyndegfree}),
gives a total of {\em six dynamical degrees of freedom\/}.

\subsection{Familiar solutions as invariant submanifolds}
\lb{subsec:fam}
The $G_{2}$ cosmologies contain a rich variety of familiar classes
of solutions as special cases. In this subsection we indicate how
these classes of solutions arise as invariant submanifolds in the
scale-invariant dynamical state space. In each case the related
equation system can be obtained by specialising the general
scale-invariant equation system and making an appropriate choice of
gauge.

\subsubsection{Vacuum $G_{2}$ cosmologies}
These solutions are described by the subset
\be
0 = \Om \ ,
\ee
which is invariant since $\Om = 0$ implies $\ptl_{t}\Om = 0$ by
Eq. (\r{dlmudot}). If the spacetime gradient $\nabla_{a}\ca$ is
timelike, one can use the timelike area gauge. The resulting
unconstrained evolution system, which has FOSH format, is given in
subsection \r{subsec:gst} [Eqs. (\r{goweq1}) -- (\r{goweq3})]. If
the spatial topology is ${\mathbb T}^{3}$, these equations describe
the {\em Gowdy vacuum spacetimes\/} that can contain gravitational
radiation with two polarisation states \ct{gow71,gow74}.

It should be noted that the evolution equation (\r{dlvdot}) for $v$
is singular on the vacuum boundary $\Om = 0$ due to the fact that
if one solves for $\ptl_{t}v$, one obtains the singular term
$\ptl_{x}\Om/\Om$. This fact means that care has to be taken in
taking limits as $\Om \rightarrow 0$, unless one is working in the
fluid-comoving gauge, in which case this problem does not arise.

\subsubsection{Diagonal $G_{2}$ cosmologies}
The orthogonally transitive $G_{2}$ cosmologies have in general
four dynamical degrees of freedom in the gravitational field, of
two different polarisation states, that are associated with the
null characteristic eigenfields $(\Sigm\pm\Nc)$ and
$(\Sigc\mp\Nm)$. With the spatial gauge choice (\r{sgfix}), it
follows that the conditions
\be
0 = \Sigc = \Nm
\ee
define an invariant submanifold which corresponds to $G_{2}$
cosmologies with one possible polarisation state only. We shall
refer to this class of solutions as {\em diagonal $G_{2}$
cosmologies\/}, because for them the line element can be written in
diagonal form (since both Killing vector fields are hypersurface
orthogonal; cf. WE \ct{waiell97}).

\subsubsection{Plane symmetrical $G_{2}$ cosmologies}
Specialising further, we can eliminate both polarisation states by
considering the invariant submanifold
\be
0 = \Sigc = \Nm = \Sigm = \Nc \ ,
\ee
which describes the class of {\em plane symmetrical $G_{2}$
cosmologies\/} (the isometry group here is a $G_{3}$ acting
multiply-transitively on flat spacelike 2-surfaces). These
solutions are the plane symmetrical analogues of the well-known
spherically symmetrical Lema\^{\i}tre--Tolman--Bondi models, in
general with non-zero fluid pressure (see Stewart and Ellis
\ct{steell68} and Eardley {\em et al\/} \ct{earetal72}). When $\gam =
1$ the evolution system reduces to a set of ODE.

\subsubsection{Self-similar $G_{2}$ cosmologies}
It is of interest to consider the $G_{2}$ cosmologies that
correspond to the {\em equilibrium points\/} (i.e., fixed points)
of the evolution system (\r{ovevol}), that are defined by the
condition
\be
\lb{selfsim}
0 = \parb_{0}{\bf X} \ .
\ee
This condition means that the dynamical state vector ${\bf X}$ is
constant on those timelike 3-surfaces whose spacelike normal
congruence is $\p_{1}$. It follows that these 3-surfaces are the
orbits of a 3-parameter homothety group $H_{3}$, i.e., solutions of
this kind are self-similar. It is important to note that the
condition (\r{selfsim}) should be imposed {\em before\/} specifying
the temporal gauge, since it uniquely fixes the gauge by specifying
the timelike frame vector field $\p_{0}$. Indeed, Eq. (\r{selfsim})
implies that both the separable area gauge condition (\r{areagc})
and the null cone gauge condition (\r{ncgc}) are satisfied.

Under these conditions, the evolution system (\r{ovevol}) reduces
to a set of ODE that govern the spatial dependence of the
models. In other words, the condition (\r{selfsim}) defines a
finite-dimensional submanifold of the infinite-dimensional
dynamical state space. Specialising further, the condition $v = 0$
defines a smaller invariant set of solutions consisting of
self-similar models whose fluid 4-velocity field $\ti{\bf u}$ is
tangent to the $H_{3}$--orbits. These solutions, which we shall
refer to as {\em fluid-aligned self-similar $G_{2}$ cosmologies\/},
have been analysed qualitatively in some detail by Hewitt {\em et
al\/} \ct{hewetal88,hewetal91,hew97}. They are of interest as
potential future asymptotic states for more general $G_{2}$
cosmologies.

\subsubsection{Spatially homogeneous $G_{2}$ cosmologies}
The conditions
\be
\lb{shcon}
0 = \parb_{1}{\bf X} \ , \hsp5 0 = \Udot \ ,
\ee
define a finite-dimensional invariant submanifold of the
infinite-dimensional dynamical state space corresponding to {\em
$G_{2}$ SH models\/}, which admit a $G_{3}$ isometry group acting
transitively on the spacelike 3-surfaces orthogonal to
$\p_{0}$.\footnote{All perfect fluid SH cosmologies of Bianchi
Type--I to Type--VII$_{h}$, apart from the exceptional
Type--VI$_{-1/9}$, admit an Abelian $G_{2}$ subgroup which acts
orthogonally transitively, and are hence included.} As with
Eq. (\r{selfsim}), the condition (\r{shcon}) should be imposed {\em
before\/} specifying the temporal gauge, since it likewise fixes
the gauge by specifying the timelike frame vector field
$\p_{0}$. Indeed, Eq. (\r{shcon}) implies that both the separable
area gauge condition (\r{areagc}) and the null cone gauge condition
(\r{ncgc}) are satisfied.

Under these conditions, the evolution system (\r{ovevol}) reduces
to a set of ODE which determines the dynamical evolution of the
models. Since $\parb_{1}{\bf X} = 0$, the $E_{1}{}^{1}$--equation
(\r{dle11dot}) decouples from the full system. Thus, if one is only
interested in the evolution of $G_{2}$ SH models, all relevant
information is given by the remaining equations, which are
analogous to the equation systems studied in WE, but with
$H$-normalisation replaced by $\beta$-normalisation. However, since
we are interested in how the $G_{2}$ SH models are related to the
$G_{2}$ cosmologies, it is necessary to retain the
$E_{1}{}^{1}$--equation. Specialising further, the condition $v =
0$ defines a smaller invariant set of solutions, the so-called {\em
non-tilted SH models\/}, in which the fluid 4-velocity field
$\ti{\bf u}$ is orthogonal to the $G_{3}$--orbits.

The equilibrium points of SH dynamics, i.e., cosmologies that admit
an $H_{4}$ acting transitively on spacetime, play an important
r\^{o}le in the $G_{2}$ dynamical state space. There are two main
subclasses. Firstly, those equilibrium points that satisfy
$(q+3\Sigp) \neq 0$ must satisfy $E_{1}{}^{1} = 0$, on account of
Eq. (\r{dle11dot}). They are thus constrained to lie in the
unphysical boundary $E_{1}{}^{1} = 0$ (see subsection
\r{subsec:unphysb}), and hence can potentially affect the $G_{2}$
dynamics near the cosmological initial singularity. The most
important examples are the {\em Kasner equilibrium set\/} (see
subsections \r{subsec:unphysb} and \r{subsec:kasnlin}) and the {\em
flat FL equilibrium point\/} (see subsection
\r{subsec:isoini}). Secondly, those SH equilibrium points that
satisfy $(q+3\Sigp) = 0$ lie in the physical part of the dynamical
state space and hence can potentially affect the evolution at late
times. The most important of these are the so-called {\em
plane-wave equilibrium points\/} (see WE \ct{waiell97}, Ch. 9).

In the present formulation it is possible to solve globally for
$\Sigp$ from the Gau\ss\ constraint equation (\r{fried}). So, for
example, in the vacuum case one thus automatically obtains a
reduced dynamical system whose dimension is equal to the number of
dynamical degrees of freedom.\footnote{Thus there exists no
drawback with the orthonormal frame approach as recently indicated
by Szyd\l owski and Demaret \ct{szydem99} (of course, a
$\beta$-normalisation can also be done in the Type--VIII and
Type--IX cases).}

It is worth noting that if one introduces the standard
Fermi-propagated diagonal frame for SH models of class A, which is
{\em not\/} the default frame choice in our formulation, the Kasner
set is represented by a {\em parabola\/} given by $2\Sigp +
\Sigm^{2} = 1$, while the Type--II vacuum solutions are {\em
straight lines\/}.

\subsection{Unphysical boundary}
\lb{subsec:unphysb}
The evolution equation (\r{dle11dot}) for the scale-invariant frame
variable $E_{1}{}^{1}$ shows that the set $E_{1}{}^{1} = 0$ defines
an invariant submanifold in an arbitrary gauge. This invariant
submanifold divides the dynamical state space into two disjoint
invariant submanifolds given by $E_{1}{}^{1} > 0$ and $E_{1}{}^{1}
< 0$. The full scale-invariant equation system is, however,
invariant under the discrete symmetry
\be
(x,\,E_{1}{}^{1}) \hsp5 \longrightarrow \hsp5
(-\,x, \,-\,E_{1}{}^{1}) \ .
\ee
The two invariant submanifolds are thus physically equivalent, and
without loss of generality we can restrict our considerations to
the case
\be
E_{1}{}^{1} > 0 \ .
\ee
The set $E_{1}{}^{1} = 0$ corresponds to unphysical states for
which the area expansion rate $\beta$ diverges ($\beta \rightarrow
+\,\infty$), typically leading to a spacetime singularity. We shall
refer to the invariant submanifold $E_{1}{}^{1} = 0$ as the {\em
unphysical boundary\/} of the infinite-dimensional dynamical state
space. It is significant that the evolution system is well-defined
on the unphysical boundary $E_{1}{}^{1} = 0$. Indeed
Eq. (\r{ovevol}) reduces to
\be
\lb{ovevolode}
{\bf A}({\bf X})\,\parb_{0}{\bf X}
= {\bf F}({\bf X}) \ ,
\ee
i.e., a system of ODE, on the unphysical boundary. It is important
to note that the {\em solutions of Eq. (\r{ovevolode})\/}, regarded
as solutions of the full evolution system (\r{ovevol}), {\em have
arbitrary $x$-dependence\/}.\footnote{An important example of such
a solution is a Kasner metric whose Kasner exponents, instead of
being constants, depend on the local coordinate $x$.} These
solutions thus do not, in general, correspond to solutions of the
EFE, and in this sense they are unphysical. Nevertheless, they do
play a significant r\^{o}le in the evolution of $G_{2}$
cosmologies. The key point is that if an orbit in the physical part
of the dynamical state space with $E_{1}{}^{1} > 0$ approaches the
unphysical boundary as $t \rightarrow -\,\infty$, then it will
shadow orbits in this boundary, i.e., the dynamics in the
unphysical boundary will determine the asymptotic dynamics of
$G_{2}$ cosmologies that are solutions to the EFE and are thus
regarded physical. In the unphysical boundary there is a hierarchy
of invariant submanifolds that influence the asymptotic dynamics of
$G_{2}$ cosmologies. Firstly note that on the unphysical boundary
the gauge fixing condition (\r{dlgsfr}) reduces to $(r-\Udot) = 0$,
which, in conjunction with the integrability condition
(\r{betaint}), implies that
\be
\lb{unphbudotdot}
\cn^{-1}\,\ptl_{t}\Udot = (q+3\Sigp)\,\Udot \ .
\ee
It follows that, in an arbitrary gauge, the condition $\Udot = 0$
defines an invariant submanifold in the unphysical boundary. The
{\em evolution equations for the invariant submanifold $0 =
E_{1}{}^{1} = \Udot$ are precisely the evolution equations for SH
models\/}, as follows from subsection \r{subsec:fam}.

Within the invariant submanifold $0 = E_{1}{}^{1} = \Udot$, the
vacuum subset $\Om = 0$ is invariant, and within this set is the
Kasner invariant set, defined by
\be
\lb{kasset}
0 = A = \Nc = \Nm \ .
\ee
These conditions imply, on account of the Gau\ss\ constraint
equation (\r{fried}) and Eqs. (\r{3rscl}) and (\r{q3Sigp}), that
\be
(q+3\Sigp) = 2 \ ,
\ee
and
\be
2\Sigp + \Sigm^{2} + \Sigc^{2} = 1 \ .
\ee
The remaining evolution equations are
\bea
\cn^{-1}\,\ptl_{t}\Sigp & = & 0 \\
\cn^{-1}\,\ptl_{t}\Sigm & = & 2\sqrt{3}\,\Sigc^{2} \\
\cn^{-1}\,\ptl_{t}\Sigc & = & -\,2\sqrt{3}\,\Sigm\,\Sigc \ ,
\eea
where the dependence of $\Sigp$, $\Sigm$ and $\Sigc$ on the local
coordinate $x$ is unrestricted. The {\em Kasner equilibrium
points\/} are given by
\be
\lb{kaspar}
\Sigc = 0 \ , \hsp5 2\Sigp + \Sigm^{2} = 1 \ ,
\ee
where $\Sigm$ is an arbitrary function of $x$. These conditions
define a {\em Kasner parabola\/} ${\cal K}$. Intuitively speaking,
the orbits in the Kasner invariant set, including the Kasner
equilibrium points, describe a $G_{2}$ cosmology with the evolution
of a Kasner vacuum solution of the EFE, but with unrestricted
$x$-dependence.

\subsection{Timelike area gauge}
\lb{subsec:gst}
In this subsection we give the evolution system in the timelike
area gauge, which was introduced in subsection
\r{subsubsec:areag}. For simplicity, we assume that the
cosmological constant is zero, $\Oml = 0$. This gauge has the
advantage of leading to an {\em unconstrained\/} evolution system
in FOSH format, as follows.
\bea
\lb{tlareae11dot}
\cn_{0}^{-1}\,\ptl_{t}E_{1}{}^{1}
& = & (q+3\Sigp)\,E_{1}{}^{1} \\
\lb{tlareasigmdot}
\cn_{0}^{-1}\,\ptl_{t}\Sigm + E_{1}{}^{1}\,\ptl_{x}\Nc
& = & (q+3\Sigp-2)\,\Sigm + 2\sqrt{3}\,\Sigc^{2}
- 2\sqrt{3}\,\Nm^{2} \\
\lb{tlareancdot}
\cn_{0}^{-1}\,\ptl_{t}\Nc + E_{1}{}^{1}\,\ptl_{x}\Sigm
& = & (q+3\Sigp)\,\Nc \\
\lb{tlareasigcdot}
\cn_{0}^{-1}\,\ptl_{t}\Sigc - E_{1}{}^{1}\,\ptl_{x}\Nm
& = & (q+3\Sigp-2-2\sqrt{3}\Sigm)\,\Sigc - 2\sqrt{3}\,\Nc\,\Nm \\
\lb{tlareanmdot}
\cn_{0}^{-1}\,\ptl_{t}\Nm - E_{1}{}^{1}\,\ptl_{x}\Sigc
& = & (q+3\Sigp+2\sqrt{3}\Sigm)\,\Nm + 2\sqrt{3}\,\Sigc\,\Nc
\eea
\bea
\lb{tlareaomdot}
\frac{f_{1}}{\Om}\,(\cn_{0}^{-1}\,\ptl_{t}
+ \frac{\gam}{G_{+}}\,v\,E_{1}{}^{1}\,\ptl_{x})\,\Om
+ f_{2}\,E_{1}{}^{1}\,\ptl_{x}v
& = & 2\,\frac{\gam}{G_{+}}\,f_{1} \\
& & \times \ [\ \frac{G_{+}}{\gam}\,(q+1)
- \sfrac{1}{2}\,(1-3\Sigp)\,(1+v^{2}) - 1\ ] \nonumber \\
\lb{tlareavdot}
\frac{f_{2}}{f_{1}}\,\Om\,(\cn_{0}^{-1}\,\ptl_{t}
- \frac{f_{3}}{G_{+}G_{-}}\,v\,E_{1}{}^{1}\,\ptl_{x})\,v
+ f_{2}\,E_{1}{}^{1}\,\ptl_{x}\Om
& = & -\,\frac{f_{2}}{f_{1}G_{-}}\,\Om\,(1-v^{2})\,
[\ \frac{(2-\gam)}{\gam}\,G_{+}\,\Udot \nonumber \\
& & \hsp5 + \ (2-\gam)\,(1-3\Sigp)\,v
- 2\,(\gam-1)\,v\ ] \ .
\eea
where
\be
\lb{ca:q3sigp}
(q+3\Sigp) =  2 - \sfrac{3}{2}\,\frac{(2-\gamma)}{G_{+}}\,
(1-v^{2})\,\Om \ .
\ee
The auxiliary variables $\Sigp$ and $\Udot$ are obtained from the
Gau\ss\ constraint equation (\r{fried}) and Eqs. (\r{areagc}) and
(\r{codac1}) as
\bea
\lb{ca:sigpt}
\Sigp & = & \sfrac{1}{2}\,(1-\Sigm^{2}-\Nc^{2}-\Sigc^{2}
-\Nm^{2}-\Om) \\
\lb{ca:udot}
\Udot & = & r \ = \  -\,3\,(\Nc\,\Sigm-\Nm\,\Sigc)
- \sfrac{3}{2}\,\frac{\gam}{G_{+}}\,\Om\,v \ .
\eea
Note that in the present case we have from Eqs. (\r{ca:q3sigp}) and
(\r{ca:sigpt}) that $q \geq \sfrac{1}{2}$, which, on account of
Eq. (\r{betaq}), guarantees that $\beta$ is a {\em monotone
function\/}.

\subsubsection{Gowdy vacuum spacetimes}
The vacuum subcase of Eqs. (\r{tlareae11dot}) -- (\r{ca:q3sigp})
describes amongst others the Gowdy spacetimes with spatial topology
${\mathbb T}^{3}$ \ct{gow71,gow74}. In characteristic normal form
these equations can be written as\footnote{On the characteristic
normal form of a FOSH evolution system, see Ref.~\ct{couhil62}.}
\bea
\lb{goweq1}
\cn_{0}^{-1}\,\ptl_{t}E_{1}{}^{1}
& = & 2E_{1}{}^{1} \\
\lb{goweq2}
(\cn_{0}^{-1}\,\ptl_{t}\pm E_{1}{}^{1}\,\ptl_{x})\,
(\Sigm\pm\Nc)
& = & (\Sigm\pm\Nc) - (\Sigm\mp\Nc)
+ 2\sqrt{3}\,(\Sigc\mp\Nm)\,(\Sigc\pm\Nm) \\
\lb{goweq3}
(\cn_{0}^{-1}\,\ptl_{t}\pm E_{1}{}^{1}\,\ptl_{x})\,
(\Sigc\mp\Nm)
& = & (\Sigc\mp\Nm) - [\,1+2\sqrt{3}\,(\Sigm\pm\Nc)\,]\,
(\Sigc\pm\Nm) \ .
\eea
Note that in the present case we can use a reparametrisation
(\r{crepara}) of $x$ to set $\ptl_{x}E_{1}{}^{1} = 0$.  We use this
representation to exemplify the following three aspects, which hold
for the whole class of $G_{2}$ cosmologies and, in suitably
generalised form, indeed for any general cosmological model.
\begin{itemize}
\item[(i)]
As the source terms on the RHS of the gravitational field equations
(\r{goweq2}) and (\r{goweq3}) (and $E_{1}{}^{1}$) must be {\em
continuous\/} for the PDE system to be well-defined in the ordinary
sense, so are the four characteristic first derivatives on the
LHS. This implies that there are four {\em unrestricted\/} first
derivatives given by $(\cn_{0}^{-1}\,\ptl_{t}\mp
E_{1}{}^{1}\,\ptl_{x})\, (\Sigm\pm\Nc)$ and
$(\cn_{0}^{-1}\,\ptl_{t}\mp E_{1}{}^{1}\,\ptl_{x})\,
(\Sigc\mp\Nm)$, which can be thus interpreted as the {\em arbitrary
information\/} (four free real-valued functions) that gravitational
radiation can propagate.
\item[(ii)]
Spacetimes with $0 = (\Sigc-\Nm) = (\Sigc+\Nm)$ {\em do\/} form an
invariant submanifold of the dynamical state space; as mentioned
before they correspond to the diagonal subcase for which the
dynamical degrees of freedom in the gravitational field are ``{\em
polarised\/}''.
\item[(iii)]
Right-propagating and left-propagating characteristic eigenfields
of the gravitational field {\em do not\/} form invariant
submanifolds of the dynamical state space; they {\em cannot\/} in
general be separated from each other. Referring to the Weyl
curvature components listed in the appendix, this reflects the fact
that a typical $G_{2}$ cosmology (and any general cosmological
model) is of algebraic Petrov type~I.
\end{itemize}
%

\section{Past asymptotics and the past attractor}
\lb{sec:past}
In this section we discuss the asymptotic evolution of the class of
orthogonally transitive $G_{2}$ cosmologies near the cosmological
initial singularity. We present evidence to support the claim that
the past attractor lies on the unphysical boundary, and is a subset
of the Kasner parabola ${\cal K}$. This analysis leads to a
discussion of the notion of {\em asymptotic silence\/}, and the
related conjecture BKL II. Because of orthogonal transitivity of
the $G_{2}$ isometry group, we do not expect to find a past
attractor of oscillatory nature. We also linearise the evolution
equations about the flat FL equilibrium point, and relate the
results to the concept of an isotropic initial singularity.

\subsection{Past attraction to the unphysical boundary}
We use the timelike area gauge characterised by Eqs. (\r{areagc}),
(\r{areagn}) and (\r{tlareagc}), and the associated evolution
system (\r{tlareae11dot}) -- (\r{ca:q3sigp}). The evolution
equation (\r{tlareae11dot}) for $E_{1}{}^{1}$, in conjunction with
Eq. (\r{ca:q3sigp}), reads
\be
\lb{paste11dot}
\cn_{0}^{-1}\,\ptl_{t}E_{1}{}^{1}
= \left[\ 2 - \sfrac{3}{2}\,\frac{(2-\gamma)}{G_{+}}\,
(1-v^{2})\,\Om\ \right]\,E_{1}{}^{1} \ .
\ee
For vacuum models, i.e., $\Om = 0$, we can solve this ODE,
obtaining $E_{1}{}^{1} = b(x)\,\exp(2\cn_{0}t)$, which implies
\be
\lb{pastreq1}
\lim_{t \rightarrow -\infty}E_{1}{}^{1} = 0 \ .
\ee
This equation will also hold for perfect fluid models which satisfy
the requirement\footnote{The physical interpretation of this
requirement is that matter does not affect the dynamics near the
cosmological initial singularity. According to conjecture BKL I
this condition will be satisfied except for special classes of
models.}
\be
\lb{pastreq2}
\lim_{t \rightarrow -\infty}\Om = 0 \ .
\ee
If the orbit of a $G_{2}$ cosmology satisfies Eq. (\r{pastreq1}),
the orbit will approach the unphysical boundary, and then we expect
that it will shadow orbits in the boundary, which are described by
the system of ODE obtained from Eq. (\r{ovevol}) by setting
$E_{1}{}^{1} = 0$. On the unphysical boundary $\Udot$ satisfies the
evolution equation (\r{unphbudotdot}), which is of the same form as
Eq. (\r{paste11dot}). It follows that along orbits in the
unphysical boundary $\lim_{t \rightarrow -\infty} \Udot = 0$, and
so we expect that typical orbits will approach the invariant
submanifold $0 = E_{1}{}^{1} = \Udot$. As mentioned in subsection
\r{subsec:unphysb}, the evolution system in this invariant
submanifold is precisely the evolution system for SH models. We
thus expect that SH dynamics will approximate $G_{2}$ dynamics
asymptotically as $t \rightarrow -\,\infty$. These heuristic
considerations suggest that we should consider the Kasner parabola
${\cal K}$ in order to localise a possible past attractor.

\subsection{Linearisation about the Kasner equilibrium set}
\lb{subsec:kasnlin}
In this subsection we perform a linearisation of the evolution
system about the Kasner equilibrium points that form the parabola
${\cal K}$, given by Eqs. (\r{kasset}) and (\r{kaspar}). We thus
linearise Eqs. (\r{tlareae11dot}) -- (\r{ca:q3sigp}) about the
values 
\be
\lb{kasequil}
\Sigm = {}_{0}\!\Sigm(x) \ , \hsp5
0 = E_{1}{}^{1} = \Nc = \Sigc = \Nm = \Om = v \ ,
\ee
where ${}_{0}\!\Sigm(x)$ is an arbitrary real-valued function of
$x$. We have to treat the evolution equation (\r{tlareavdot}) for
$v$ in a special manner, due to the fact that the term
$E_{1}{}^{1}\,\ptl_{x}\Om/\Om$ is singular on ${\cal K}$. We first
linearise Eq. (\r{tlareaomdot}) for $\Om$ and then use the solution
of this equation to show that the singular term in
Eq. (\r{tlareavdot}) can be neglected when linearising it.
Rescaling the time variable so that $\cn_{0} = 1$, we then obtain
the following {\em system of linear ODE\/} in $t$, with
$x$-dependent coefficient ${}_{0}\!\Sigm$:
\bea
\lb{linkaseq1}
\ptl_{t}E_{1}{}^{1}
& = & 2E_{1}{}^{1} \\
\ptl_{t}\Sigm
& = & -\,\sfrac{3}{2}\,(2-\gam)\,{}_{0}\!\Sigm\,\Om \\
\ptl_{t}\Nc
& = & 2\Nc - (\ptl_{x}{}_{0}\!\Sigm)\,E_{1}{}^{1} \\
\ptl_{t}\Sigc
& = & -\,2\sqrt{3}\,{}_{0}\!\Sigm\,\Sigc \\
\ptl_{t}\Nm
& = & 2\,(1+\sqrt{3}{}_{0}\!\Sigm)\,\Nm \\
\ptl_{t}\Om
& = & \sfrac{3}{2}\,(2-\gam)\,(1+{}_{0}\!\Sigm^{2})\,\Om \\
\lb{linkaseq7}
\ptl_{t}v
& = & \sfrac{1}{2}\,[\,3\gam-2-3\,(2-\gam)\,{}_{0}\!\Sigm^{2}\,]\,v
+ 3\,\frac{(2-\gam)}{\gam}\,{}_{0}\!\Sigm\,\Nc \ .
\eea
The general solution of Eq. (\r{linkaseq1}) is
\be
\lb{linkassol0}
E_{1}{}^{1} = b(x)\,\e^{2t} \ .
\ee
We can use a reparametrisation (\r{crepara}) of $x$ to set $b(x) =
1$. The resulting general solution of the linear ODE system is then
given by (including the zeroth-order contribution to $\Sigm$)
\bea
\lb{linkassol1}
\Sigm & = & -\,\sfrac{1}{\sqrt{3}}\,k(x)\left[\,
1 - \frac{\Om}{1+\sfrac{1}{3}k^{2}(x)}\,\right] \\
\lb{linkassol2}
\Nc & = & [\,a_{2}(x)+\sfrac{1}{\sqrt{3}}\,t\,
\ptl_{x}k(x)\,]\,\e^{2t} \\
\lb{linkassol3}
\Sigc & = & a_{3}(x)\,\e^{2k(x)t} \\
\lb{linkassol4}
\Nm & = & a_{4}(x)\,\e^{2[1-k(x)]t} \\
\lb{linkassol5}
\Om & = & a_{5}(x)\,
\e^{\sfrac{3}{2}(2-\gam)[1+\sfrac{1}{3}k^{2}(x)]t} \\
\lb{linkassol6}
v & = &  a_{6}(x)\,
\e^{\sfrac{1}{2}[3\gam-2-(2-\gam)k^{2}(x)]t}
- \sfrac{2}{\sqrt{3}}\,\frac{k(x)}{\gam\,[1+\sfrac{1}{3}k^{2}(x)]}
\,\e^{2t} \nonumber \\
& & \hsp5 \times\left[\,a_{2}(x) + \sfrac{1}{\sqrt{3}}\,
\ptl_{x}k(x)\,\left(t-\frac{2}{3(2-\gam)
[1+\sfrac{1}{3}k^{2}(x)]}\right)\,\right] \ .
\eea
For convenience, and to agree with the notation of Rendall and
Weaver \ct{renwea2001}, we write
\be
{}_{0}\!\Sigm(x) = -\,\sfrac{1}{\sqrt{3}}\,k(x)
\ee
for the limiting value of $\Sigm$ as $t \rightarrow -\,\infty$. The
solution of the linear ODE system suggests that an arc ${\cal
K}_{A}$ of the Kasner equilibrium set ${\cal K}$ attracts
neighbouring orbits (i.e., is a local attractor). This arc is
defined by the requirement that in each exponential function in the
solution (\r{linkassol1}) -- (\r{linkassol6}) the independent
variable $t$ has a positive coefficient, so that the solution
approaches the equilibrium point (\r{kasequil}) as $t \rightarrow
-\,\infty$. The size of the arc depends on whether the model is
polarised (i.e., diagonal) or not, and on the equation of state
parameter $\gam$ of the fluid, as shown in Tab.~\r{tab3}.
\begin{table}[!htb]
\begin{center}
\begin{tabular}{lc} 
\hline \hline
Class of models & Attracting arc ${\cal K}_{A}$ \\
\hline
Vacuum/polarised & all of ${\cal K}$ \\
Fluid/polarised  &
$-\,\frac{(3\gam-2)^{1/2}}{\sqrt{3}(2-\gam)^{1/2}}
< {}_{0}\!\Sigm < 0$ \\
Vacuum, or fluid with $1 \leq \gam < 2$/unpolarised &
$-\,\sfrac{1}{\sqrt{3}} \leq {}_{0}\!\Sigm < 0$ \\
\hline \hline
\end{tabular}
\end{center}
\caption{Attracting arc ${\cal K}_{A}$ on the Kasner parabola
${\cal K}$}
\lb{tab3}
\end{table}
We stress that this linear analysis does not prove that ${\cal
K}_{A}$ is a local attractor.

Over the past 11 years a number of rigorous analyses of the past
asymptotic behaviour of $G_{2}$ cosmologies have been given which
enable us to make precise statements about ${\cal K}_{A}$. Firstly,
Isenberg and Moncrief \ct{isemon90} have proved that every
polarised Gowdy vacuum solution with spatial topology ${\mathbb
T}^{3}$ is past asymptotic to a Kasner solution, showing that {\em
the Kasner equilibrium set ${\cal K}$ is the global past attractor
for this class of models\/}. Secondly, Kichenassamy and Rendall
\ct{kicren98} used an analysis based on the Fuchsian
algorithm to prove that a general family\footnote{That is, a family
whose initial data depends on four arbitrary real-valued
functions.} of unpolarised Gowdy vacuum solutions with spatial
topology ${\mathbb T}^{3}$ is past asymptotic to the arc ${\cal
K}_{A}$, as given in Tab.~\r{tab3}. In a recent development,
Rendall \ct{ren2001} has used the $\beta$-normalised
scale-invariant FOSH evolution system for Gowdy vacuum spacetimes
to argue that the arc ${\cal K}_{A}$ is in fact a local past
attractor in the unpolarised case, for models which satisfy the
so-called ``low velocity'' condition $0 \leq v_{\rm Gowdy} < 1$,
where the Gowdy ``velocity parameter'' corresponds to
\be
v_{\rm Gowdy} =
\sqrt{3}\,(\Sigm^{2}+\Sigc^{2})^{1/2} \ ,
\ee
and thus quantifies the magnitude of the {\em transverse\/} shear
rate of the timelike reference congruence $\p_{0}$. It is known,
however, that the arc ${\cal K}_{A}$ is not the global past
attractor for Gowdy vacuum spacetimes since solutions which develop
so-called spikes violate the inequality $0 \leq v_{\rm Gowdy} < 1$
at those points at which a spike occurs (see Rendall and Weaver
\ct{renwea2001}, Berger and Moncrief \ct{bermon93} and Hern and
Stewart \ct{herste98}). Finally, Anguige \ct{ang2000b} has proved
that a general family of diagonal $G_{2}$ cosmologies with a
perfect fluid matter source is past asymptotic to the arc ${\cal
K}_{A}$ as given in Tab.~\r{tab3}.

It also follows from Refs. \ct{isemon90}, \ct{kicren98} and
\ct{ang2000b} that the solution (\r{linkassol0}) --
(\r{linkassol6}) to the linear equations gives the correct past
asymptotic form of a general class of solutions in a neighbourhood
of the local attractor ${\cal K}_{A}$ for polarised and unpolarised
Gowdy vacuum spacetimes and for diagonal $G_{2}$ cosmologies with a
perfect fluid matter source. We anticipate that it will also do so
for orthogonally transitive perfect fluid $G_{2}$ cosmologies, but
this remains to be proven.

Finally, we note that if the restriction ${}_{0}\!\Sigm >
-\,\frac{(3\gam-2)^{1/2}}{\sqrt{3}(2-\gam)^{1/2}}$, which arises in
the polarised perfect fluid $G_{2}$ case in Tab.~\r{tab3}, does not
hold, then the peculiar velocity $v$ of the fluid will remain
significant as $t \rightarrow -\,\infty$, hinting at the existence
of another local attractor distinct from ${\cal K}_{A}$. Experience
with SH cosmologies (see, e.g., Ref.~\ct{hewwai92}) suggests that
$v$ will approach its extreme values, i.e.,
$\lim_{t\rightarrow -\infty}v = \pm\,1$. This matter requires further
investigation.

\subsection{Asymptotic silence}
We now give a brief discussion of the conjecture BKL II, in the
light of the previous two subsections. This conjecture is part of
the folklore of mathematical cosmology and does not have a precise
statement. We can best explain the essence of the conjecture by
quoting from BKL \ct{bkl82}, p656:
\begin{quotation}
``\dots in the asymptotic vicinity of the singular point the
Einstein equations are effectively reduced to a system of ordinary
differential equations with respect to time: the spatial
derivatives enter these equations `passively' without influencing
the character of the solution.''
\end{quotation}
Another way of expressing the idea heuristically is to say that the
evolution at different spatial points decouples near the
cosmological initial singularity. The FOSH format of the evolution
system that we have given, namely
\be
\lb{pdeevol}
{\bf A}\,\ptl_{t}{\bf X}
+ {\bf B}\,E_{1}{}^{1}\,\ptl_{x}{\bf X}
= {\bf F}({\bf X}) \ ,
\ee
(using the timelike area gauge with $\cn_{0} = 1$), sheds light on
this idea of spatial decoupling, since we have shown that $\lim_{t
\rightarrow -\infty} E_{1}{}^{1} = 0$ when $\lim_{t \rightarrow
-\infty}\Om = 0$. This result means that as one follows a timeline
into the past towards the cosmological initial singularity, the
local null cone (and hence the local fluid sound cone embedded
therein) collapses onto the timeline, showing that geometrical
information propagation between neighbouring timelines is
asymptotically eliminated, as illustrated in Fig.~\r{fig1}.
\begin{figure}[!htb]
\vspace*{0.3cm}
\epsfxsize=20pc\epsfbox{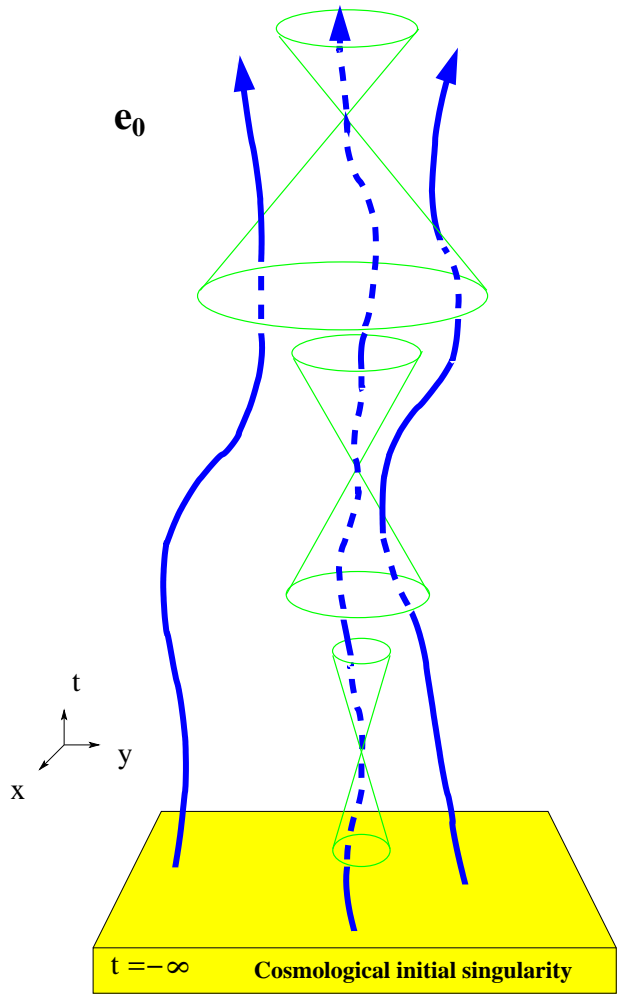}
\vspace*{0.5cm}
\caption{Schematic representation of the phenomenon of
``asymptotic silence'' of the gravitational field dynamics in the
approach of the cosmological initial singularity.}
\lb{fig1}
\end{figure}
We shall refer to this phenomenon as ``{\em asymptotic silence\/}''
of the gravitational field dynamics as the cosmological initial
singularity is approached.\footnote{On the original idea of
``silent cosmological models'', see Matarrese {\em et al\/}
\ct{matetal94}; on their dynamical consequences,
Ref.~\ct{hveetal97}.} As regards the quotation from BKL \ct{bkl82},
there are two ways of reducing the evolution system of PDE to a
system of ODE. Firstly, one can set $E_{1}{}^{1} = 0$ in
Eq. (\r{pdeevol}), obtaining the system of ODE that describes the
dynamics on the unphysical boundary:\footnote{This is related to
the so-called ``velocity-dominated'' system obtained by dropping
the spatial derivatives from the evolution system but not from the
constraint equations; see, e.g., Andersson and Rendall
\ct{andren2001}.}
\be
\lb{odeevol}
{\bf A}\,\ptl_{t}{\bf X}
= {\bf F}({\bf X}) \ .
\ee
Secondly, one can consider the system of linear ODE in the
neighbourhood of the Kasner equilibrium set, Eqs. (\r{linkaseq1})
-- (\r{linkaseq7}). We have seen that the system of linear ODE do
produce the correct past asymptotic form of the solutions near the
locally attracting Kasner arc ${\cal K}_{A}$. One can also consider
the relation between Eqs. (\r{odeevol}) and (\r{pdeevol}). In view
of the fact that $\lim_{t \rightarrow -\infty} E_{1}{}^{1} = 0$,
one might expect that the spatial derivative term ${\bf
B}\,E_{1}{}^{1}\,\ptl_{x}{\bf X}$ in Eq. (\r{pdeevol}) would be
negligible compared to the other two terms, as $t \rightarrow
-\,\infty$. Calculating each term in Eq. (\r{pdeevol}) using the
asymptotic solution (\r{linkassol0}) -- (\r{linkassol6}) shows that
the spatial derivative terms are in fact negligible asymptotically
in a neighbourhood of the Kasner arc ${\cal K}_{A}$. This property
does {\em not\/} hold at isolated points in solutions which develop
spikes, since certain partial derivatives in $\ptl_{x}{\bf X}$
become unbounded, so that the term ${\bf
B}\,E_{1}{}^{1}\,\ptl_{x}{\bf X}$ is not negligible
there. Nevertheless, the spatial derivative terms appear to act
``passively'', so that one still has silent, Kasner-like dynamics
locally.

\subsection{Isotropic initial singularities}
\lb{subsec:isoini}
Our discussion in subsection \r{subsec:kasnlin} concerns the past
asymptotic behaviour of general classes of $G_{2}$ cosmologies,
which, according to conjecture BKL I, satisfy $\lim_{t \rightarrow
-\infty}\Om = 0$. There are, however, special classes of models
that violate this condition, the most important being models with
an {\em isotropic initial singularity\/} (see Goode and Wainwright
\ct{goowai85}), which satisfy
\be
\lim_{t \rightarrow -\infty}\Om = 1 \ , \hsp5
\lim_{t \rightarrow -\infty}v = 0 \ , \hsp5
\lim_{t \rightarrow -\infty}\Sig^{2} = 0 \ .
\ee
Their evolution near the cosmological initial singularity is
approximated by the flat FL model. We can understand the generality
of the isotropic initial singularity by linearising the evolution
equations (\r{tlareae11dot}) -- (\r{ca:q3sigp}) about the flat FL
equilibrium point which lies on the unphysical boundary, and is
given by
\be
\Om = {}_{0}\!\Om = 1 \ , \hsp5
0 = E_{1}{}^{1} = \Sigm = \Nc = \Sigc = \Nm = v \ .
\ee
Setting $\cn_{0} = 1$, we thus obtain the following system of ODE:
\bea
\ptl_{t}E_{1}{}^{1}
& = & \sfrac{1}{2}\,(3\gam-2)\,E_{1}{}^{1} \\
\ptl_{t}\Sigm
& = & -\,\sfrac{3}{2}\,(2-\gam)\,\Sigm \hspace{22mm}
\ptl_{t}\Nc
\ = \ \sfrac{1}{2}\,(3\gam-2)\,\Nc \\
\ptl_{t}\Sigc
& = & -\,\sfrac{3}{2}\,(2-\gam)\,\Sigc \hspace{22mm}
\ptl_{t}\Nm
\ = \ \sfrac{1}{2}\,(3\gam-2)\,\Nm \\
\ptl_{t}\Om
& = & \sfrac{3}{2}\,(2-\gam)\,(1-\Omega) \hspace{22mm}
\ptl_{t}v
\ = \ \sfrac{1}{2}\,(3\gam-2)\,v \ .
\eea
Using a reparametrisation (\r{crepara}) of $x$,
we obtain $E_{1}{}^{1} = \exp(\sfrac{1}{2}(3\gam-2)t)$, and
\bea
\Sigm & = & a_{1}(x)\,\e^{-\sfrac{3}{2}(2-\gam)t} \hspace{22mm}
\Nc \ = \ a_{2}(x)\,\e^{\sfrac{1}{2}(3\gam-2)t} \\
\Sigc & = & a_{3}(x)\,\e^{-\sfrac{3}{2}(2-\gam)t} \hspace{22mm}
\Nm \ = \ a_{4}(x)\,\e^{\sfrac{1}{2}(3\gam-2)t} \\
\Om & = & 1 + a_{5}(x)\,\e^{-\sfrac{3}{2}(2-\gam)t} \hspace{20mm}
v \ = \ a_{6}(x)\,\e^{\sfrac{1}{2}(3\gam-2)t} \ .
\eea
Note that it follows from the present result that for a $G_{2}$
cosmology to have an isotropic initial singularity the conditions
\be
0 = a_{1}(x) = a_{3}(x) = a_{5}(x)
\ee
need to be satisfied. This amounts to setting precisely {\em
half\/} the initial data compared to the full $G_{2}$ case. For
rigorous results see Claudel and Newman \ct{clanew98} and Anguige
and Tod \ct{angtod99}.

\section{Concluding remarks and outlook}
\lb{sec:concl}
In this paper we have shown how to formulate the EFE for
orthogonally transitive $G_{2}$ cosmologies with a perfect fluid
matter source as an autonomous system of PDE with evolution
equations in FOSH format, using scale-invariant dependent
variables. As stated in the introduction, one of our goals is to
provide a flexible framework for analysing $G_{2}$ dynamics. A
potential user of this paper, someone who wishes to apply the
equation systems that we have derived to do numerical or rigorous
analyses, need not be familiar with the orthonormal frame
formalism. For such a reader the heart of the paper is the
scale-invariant equation system in section \r{subsec:dimlesseq},
the discussion of the gauge choices in section \r{sec:gauge}, and
the overview in section \r{subsec:ov}. The explicit equation system
in section \r{subsec:gst} for the timelike area gauge may also be
of use. On the other hand, for someone interested in the structure
of the space of cosmological solutions of the EFE, the relevant
sections are \r{subsec:fam}, \r{subsec:unphysb} and \r{sec:past}.

An important aspect of our formulation is that it incorporates the
SH models as a special case, thereby shedding light on how SH
dynamics influences $G_{2}$ dynamics. The $H$-normalised
scale-invariant dependent variables, when used in a dynamical
systems setting, have proved effective in all four aspects of
analysis of SH models, i.e., exact solutions, heuristic, numerical
and rigorous. In particular, the dynamical systems framework
suggested various heuristic ways of gaining insight into the SH
dynamics, using local stability arguments and the notion of
shadowing of orbits in the dynamical state space, in conjunction
with the hierarchy of Bianchi invariant submanifolds. These methods
have in turn led to proofs of various conjectures concerning SH
dynamics (e.g., Ringstr\"{o}m \ct{rin2000}) and to the prediction
of new dynamical phenomena (e.g., asymptotic self-similarity
breaking and Weyl curvature dominance; see Wainwright {\em et al\/}
\ct{waietal99}). Our initial success in describing some aspects of
the $G_{2}$ dynamics at early times in terms of a local past
attractor, as given in section \r{sec:past}, suggests that our new
formulation of the evolution for perfect fluid $G_{2}$ cosmologies
will prove equally effective. The three main problems concerning
$G_{2}$ dynamics that we intend to investigate in the future are
\begin{itemize}
\item[(i)]
the asymptotic dynamics at early times when the peculiar velocity
is dynamically significant,
\item[(ii)]
the local stability of self-similar models and the asymptotic
dynamics at late times, and
\item[(iii)]
the dynamics of $G_{2}$ cosmologies that are close to FL in some
epoch.
\end{itemize}

In conclusion, we believe that many of the ideas discussed in the
present paper, when appropriately modified, will be of relevance in
a much broader context in mathematical cosmology. For example, the
notion of an infinite-dimensional scale-invariant dynamical state
space with a hierarchical skeleton structure will be a useful guide
for exploring more general cosmological spacetimes. We anticipate
that concepts such as geometrical information propagation,
asymptotic silence, and a past attractor located on the unphysical
boundary of the dynamical state space, whose dynamics is described
by a system of ODE, will play an important r\^{o}le in clarifying
the dynamical content of the conjectures BKL I and BKL II
concerning cosmological initial singularities. Likewise, these
ingredients should be useful for studying almost-FL dynamical
states near the cosmological initial singularity or at intermediate
and late times, as well as other, more generic, aspects of these
dynamical regimes.

\section*{Acknowledgements}
We thank Woei Chet Lim and Alan Rendall for many helpful
comments. We have appreciated numerous stimulating discussions
during June/July 2001 with participants of the workshop on
``Mathematical Cosmology'', which helped clarify our understanding
of $G_{2}$ cosmologies and contributed to the final form of this
paper. This workshop was held at the Internationales Erwin
Schr\"{o}dinger Institut f\"{u}r Mathematische Physik at Wien,
Austria, whose generous hospitality is gratefully acknowledged. HvE
was in part supported by the Deutsche Forschungsgemeinschaft (DFG)
at Bonn, Germany. CU was in part supported by the Swedish Science
Council. JW was in part supported by a grant from the Natural
Sciences and Engineering Research Council of Canada. The computer
algebra packages {\tt REDUCE} and {\tt MAPLE} proved to be
invaluable tools.

\appendix
\section{Appendix}
\subsection{Connection components in terms of frame variables}
The inverse area density of the $G_{2}$--orbits is given by
\be
\ca^{-1} = e_{2}{}^{2}\,e_{3}{}^{3} - e_{2}{}^{3}\,e_{3}{}^{2} \ .
\ee
Then it follows from the dimensional commutator equations
(\r{gaugefix}) -- (\r{e3agrad}) that
\bea
\alpha & = & -\,N^{-1}\,\frac{\ptl_{t}e_{1}{}^{1}}{e_{1}{}^{1}}
\hspace{20mm}
\udot_{1} \ = \ e_{1}{}^{1}\,\frac{\ptl_{x}N}{N} \\
\beta & = & \sfrac{1}{2}\,N^{-1}\,\frac{\ptl_{t}\ca}{\ca}
\hspace{23mm}
a_{1} \ = \ -\,\sfrac{1}{2}\,e_{1}{}^{1}\,\frac{\ptl_{x}\ca}{\ca} \\
\sigm & = & -\,\sfrac{1}{2\sqrt{3}}\,\ca\,N^{-1}\,
(e_{3}{}^{3}\,\ptl_{t}e_{2}{}^{2}
- e_{2}{}^{2}\,\ptl_{t}e_{3}{}^{3}
+ e_{2}{}^{3}\,\ptl_{t}e_{3}{}^{2}
- e_{3}{}^{2}\,\ptl_{t}e_{2}{}^{3}) \\
\nc & = & \sfrac{1}{2\sqrt{3}}\,\ca\,e_{1}{}^{1}\,
(e_{3}{}^{3}\,\ptl_{x}e_{2}{}^{2}
- e_{2}{}^{2}\,\ptl_{x}e_{3}{}^{3}
+ e_{2}{}^{3}\,\ptl_{x}e_{3}{}^{2}
- e_{3}{}^{2}\,\ptl_{x}e_{2}{}^{3}) \\
\sigc & = & \sfrac{1}{2\sqrt{3}}\,\ca\,N^{-1}\,
(e_{2}{}^{3}\,\ptl_{t}e_{2}{}^{2}
- e_{2}{}^{2}\,\ptl_{t}e_{2}{}^{3}
+ e_{3}{}^{2}\,\ptl_{t}e_{3}{}^{3}
- e_{3}{}^{3}\,\ptl_{t}e_{3}{}^{2}) \\
\nm & = & \sfrac{1}{2\sqrt{3}}\,\ca\,e_{1}{}^{1}\,
(e_{2}{}^{3}\ptl_{x}e_{2}{}^{2}
- e_{2}{}^{2}\,\ptl_{x}e_{2}{}^{3}
+ e_{3}{}^{2}\,\ptl_{x}e_{3}{}^{3}
- e_{3}{}^{3}\,\ptl_{x}e_{3}{}^{2}) \\
\Om_{1} & = & \sfrac{1}{2}\,\ca\,N^{-1}\,
(e_{2}{}^{3}\,\ptl_{t}e_{2}{}^{2}
- e_{2}{}^{2}\,\ptl_{t}e_{2}{}^{3}
- e_{3}{}^{2}\,\ptl_{t}e_{3}{}^{3}
+ e_{3}{}^{3}\,\ptl_{t}e_{3}{}^{2}) \\
\np & = & -\,\sfrac{1}{2}\,\ca\,e_{1}{}^{1}\,
(e_{2}{}^{3}\,\ptl_{x}e_{2}{}^{2}
- e_{2}{}^{2}\,\ptl_{x}e_{2}{}^{3}
- e_{3}{}^{2}\,\ptl_{x}e_{3}{}^{3}
+ e_{3}{}^{3}\,\ptl_{x}e_{3}{}^{2}) \ .
\eea
%

\subsection{Scale-invariant curvature variables}
\lb{subsec:3ricweyl}
We define additional $\beta$-normalised curvature variables by
\be
(\,\ck, \,\casc_{\dots}, \,\ce_{\dots}, \,\ch_{\dots}\,)
:= (\,-\sfrac{1}{2}{}^{*}\!R, \,{}^{*}\!S_{\dots},
\,E_{\dots}, \,H_{\dots}\,)/(3\beta^{2}) \ .
\ee
Then we obtain for orthogonally transitive $G_{2}$ cosmologies with
a perfect fluid matter source the following expressions \enl

\noindent
{\em Non-zero 3-Ricci curvature variables\/}:  \nopagebreak
\bea
\ck & = & -\,\sfrac{2}{3}\,(E_{1}{}^{1}\,\ptl_{x}-r
-\sfrac{3}{2}A)\,A + (\Nc^{2}+\Nm^{2}) \\
\casc_{+} & = & -\,\sfrac{1}{9}\,(E_{1}{}^{1}\,\ptl_{x}-r)\,A
+ \sfrac{2}{3}\,(\Nc^{2}+\Nm^{2}) \\
\casc_{-} & = & \sfrac{1}{3}\,(E_{1}{}^{1}\,\ptl_{x}-r
-2A)\,N_{\times} + \sfrac{2}{3}\,N_{+}\,N_{-} \\
\casc_{\times} & = & -\,\sfrac{1}{3}\,(E_{1}{}^{1}\,\ptl_{x}-r
-2A)\,N_{-} + \sfrac{2}{3}\,N_{+}\,N_{\times} \ .
\eea

\noindent
{\em Non-zero characteristic Weyl curvature
variables\/}:\footnote{Again, we correct some sign errors in the
expressions given in Refs. \ct{hveell99} and \ct{hveetal2000}.}
\bea
\lb{dlep}
\ce_{+} & = & -\,\sfrac{1}{9}\,(E_{1}{}^{1}\,\ptl_{x}-r)\,A
+ \sfrac{2}{3}\,(\Nc^{2}+\Nm^{2})
+ \sfrac{1}{3}\,\Sigp
- \sfrac{1}{3}\,(\Sigm^{2}+\Sigc^{2})
+ \sfrac{1}{6}\,\gam\,G_{+}^{-1}\,\Om\,v^{2} \\
\lb{dlhp}
\ch_{+} & = & -\,N_{-}\,\Sigm - N_{\times}\,\Sigc \\
\lb{dlgwm1}
(\ce_{-}\pm\ch_{\times}) & = & \pm\,\sfrac{1}{3}\,(E_{1}{}^{1}\,
\ptl_{x}-r\mp3\Sigp-A)\,(\Sigm\pm\Nc) \mp \sfrac{2}{3}\,N_{+}\,
(\Sigc\mp\Nm) + \sfrac{1}{3}\,(\Sigm-A\,\Nc) \\
\lb{dlgwm2}
(\ce_{\times}\mp\ch_{-}) & = & \pm\,\sfrac{1}{3}\,(E_{1}{}^{1}\,
\ptl_{x}-r\mp3\Sigp-A)\,(\Sigc\mp\Nm) \pm \sfrac{2}{3}\,N_{+}\,
(\Sigm\pm\Nc) + \sfrac{1}{3}\,(\Sigc+A\,\Nm) \ .
\eea
%

\subsection{Line element and scale-invariant dependent variables
for area gauges}
\lb{subsec:areads2}
Introducing in the separable area gauge a line element of the form
\be
{\rm d}s^{2} = \ell_{0}^{2}\,\left[\ -\,\e^{2f(t,x)}\,{\rm d}t^{2}
+ \e^{2g(t,x)}\,{\rm d}x^{2} +
\e^{2\cn_{0}t}\,m(x)\left(\e^{P(t,x)}\, ({\rm d}y+Q(t,x){\rm
d}z)^{2} + \e^{-P(t,x)}\,{\rm d}z^{2}\right)\ \right] \ ,
\ee
the area expansion rate is given by
\be
\beta = \ell_{0}^{-1}\,\cn_{0}\,\e^{-f} \ .
\ee
Then we obtain the following expressions for the non-zero
scale-invariant dependent variables:
\bea
(\,\cn^{-1}, \,E_{1}{}^{1}\,)
& = & (\,\cn_{0}^{-1}, \,\cn_{0}^{-1}\,\e^{f-g}\,) \\
(\,\Sigp, \,\Udot\,)
& = & \left(\,\sfrac{1}{3}\,(1-\cn_{0}^{-1}\,\ptl_{t}g),
\,E_{1}{}^{1}\,\ptl_{x}f\,\right) \\
(\,\Sigm, \,\Nc\,)
& = & \sfrac{1}{2\sqrt{3}}\,(\,\cn_{0}^{-1}\,\ptl_{t}P,
\,-\,E_{1}{}^{1}\,\ptl_{x}P\,) \\
(\,\Sigc, \,\Nm\,)
& = & \sfrac{1}{2\sqrt{3}}\,\e^{P}\,(\,\cn_{0}^{-1}\,\ptl_{t}Q,
\,E_{1}{}^{1}\,\ptl_{x}Q\,) \\
\lb{separeaa}
A & = & -\,\sfrac{1}{2}\,\frac{{\rm d}\ln m(x)}{{\rm d}x}\,
E_{1}{}^{1} \ ,
\eea
and $R = -\,\sqrt{3}\,\Sigc$ and $\Np = \sqrt{3}\,\Nm$. Employing
an $x$-reparametrisation (\r{crepara}) to set $m(x) = \exp(-2D_{\rm
sa}x)$, with $D_{\rm sa}$ a constant, implies $A = D_{\rm
sa}\,E_{1}{}^{1}$. Apart from the sign of $t$, these expressions
reduce to the Gowdy vacuum line element of Rendall and Weaver
\ct{renwea2001} when $\cn_{0} =
\sfrac{1}{2}$, $m(x) = 1$, $f(t,x) =
\sfrac{1}{4}\,[\,\lambda(t,x)+3t\,]$ and $g(t,x) =
\sfrac{1}{4}\,[\,\lambda(t,x)-t\,]$. Note that then
$\ptl_{x}E_{1}{}^{1} = 0$.

\subsection{Propagation of constraint equations}
\lb{subsec:consprop}

\noindent
{\em Propagation of dimensional constraint equations\/}:
\nopagebreak
\bea
N^{-1}\,\ptl_{t}(C_{\rm com})^{A}{}_{12}
& = & -\,(\alpha+\beta+\sqrt{3}\sigm)\,(C_{\rm com})^{A}{}_{12}
- (\sqrt{3}\sigc+\Om_{1})\,(C_{\rm com})^{A}{}_{31} \nonumber \\ 
& & \hsp5 - \ e_{2}{}^{A}\,(C_{\rm Codacci})_{1} \\
N^{-1}\,\ptl_{t}(C_{\rm com})^{A}{}_{31}
& = & -\,(\alpha+\beta-\sqrt{3}\sigm)\,(C_{\rm com})^{A}{}_{31}
- (\sqrt{3}\sigc-\Om_{1})\,(C_{\rm com})^{A}{}_{12} \nonumber \\ 
& & \hsp5 - \ e_{3}{}^{A}\,(C_{\rm Codacci})_{1} \\
N^{-1}\,\ptl_{t}(C_{\rm Gau\ss})
& = & -\,(\alpha+\beta)\,(C_{\rm Gau\ss})
- 4\,(\udot_{1}+a_{1})\,(C_{\rm Codacci})_{1} \\
N^{-1}\,\ptl_{t}(C_{\rm Codacci})_{1}
& = & -\,(\alpha+3\,\beta)\,(C_{\rm Codacci})_{1}
- \sfrac{1}{4}\,(\udot_{1}-a_{1})\,(C_{\rm Gau\ss}) \\
\cn^{-1}\,\ptl_{t}(C_{\beta}) & = & -\,(1-3\Sigp)\,(C_{\beta}) \ .
\eea
For simplicity, in deriving these results we have used the gauge
fixing condition (\r{gaugefix}) as an identity. \enl

\noindent
{\em Propagation of dimensionless gauge fixing conditions\/}:
\bea
\lb{gfcpropnc}
\cn^{-1}\,\ptl_{t}({\cal C}_{\Udot})_{\rm nc}
& = & (q+1)\,({\cal C}_{\Udot})_{\rm nc} \\
\lb{gfcpropsa}
\cn_{0}^{-1}\,\ptl_{t}({\cal C}_{\Udot})_{\rm sa}
& = & (q+3\Sigp)\,({\cal C}_{\Udot})_{\rm sa} \\
\lb{gfcpropco}
\cn^{-1}\,\ptl_{t}({\cal C}_{\Udot})_{\rm fc}
& = & [\,(q+3\Sigp)+3\,(\gam-1)\,(1-\Sigp)\,]\,
({\cal C}_{\Udot})_{\rm fc} \\
\lb{gfcpropsync}
\cn^{-1}\,\ptl_{t}({\cal C}_{\Udot})_{\rm sync}
& = & (q+3\Sigp)\,({\cal C}_{\Udot})_{\rm sync} \ .
\eea
%



\end{document}